\documentclass[journal]{IEEEtran}
\onecolumn

\usepackage{amssymb}  
\usepackage{amsthm}
\usepackage{amsmath}
\usepackage{graphicx}
\usepackage{cite}
\usepackage[center]{caption}
\usepackage[font=small]{caption}
\usepackage[utf8]{inputenc}

\ifCLASSINFOpdf
 
\else
 
 \fi

\begin{document}
\pagenumbering{arabic}

\title{Moment-Based Exact Uncertainty Propagation Through Nonlinear Stochastic Autonomous Systems}

\author{Ashkan Jasour%
, Allen Wang%
, and Brian C. Williams \\
Computer Science and Artificial Intelligence Laboratory (CSAIL)\\
Massachusetts Institute of Technology (MIT)\\
\tt\small \{jasour, allenw, williams\} @ mit.edu
}


\maketitle

\begin{abstract} In this paper, we address the problem of uncertainty propagation through nonlinear stochastic dynamical systems. More precisely, given a discrete-time continuous-state probabilistic nonlinear dynamical system, we aim at finding the sequence of the moments
of the probability distributions of the system states up to any desired order over the given planning horizon. Moments of  uncertain states can  be used  in  estimation,  planning, control, and safety analysis  of  stochastic  dynamical  systems. Existing approaches to address moment propagation problems provide approximate descriptions of the moments and are mainly limited to particular set of uncertainties, e.g., Gaussian disturbances. In this paper, to describe the moments of uncertain states, we introduce trigonometric and also mixed-trigonometric-polynomial moments. Such moments allow us to obtain closed deterministic dynamical systems that describe the \textit{exact} time evolution of the moments of uncertain states of an important class of autonomous and robotic systems  including  underwater, ground, and  aerial  vehicles,  robotic  arms  and  walking robots. Such obtained deterministic dynamical systems can be used, in a receding horizon fashion, to propagate the uncertainties over the planning horizon in \textit{real-time}. To illustrate the performance of the proposed method, we benchmark our method against existing approaches including linear, unscented transformation, and sampling based uncertainty propagation methods that are widely used in estimation, prediction, planning, and control problems.

\end{abstract}

\IEEEpeerreviewmaketitle

\section{Introduction}

Uncertainty quantification plays an important role in estimation, planning, and control problems under uncertainties.
In this paper, we use moment based representation of uncertainties. Moments of uncertainties, e.g., random variables, are the generalization of mean and covariance and are defined as expected values of monomials of random variables. Moments of uncertainties can be used in estimation, planning, control, and safety analysis of uncertain dynamical systems. For example, finite sequence of the moments are used in \cite{mom_con_1, mom_con_2, mom_con_3, mom_con_4} for risk bounded control of probabilistic nonlinear and linear systems, in \cite{mom_opt_1,mom_opt_2,mom_opt_3} to solve chance-constrained optimization problems, in \cite{mom_risk_1, mom_risk_2} for nonlinear risk estimation, in \cite{mom_pdf} to compute the probability density function of uncertainties, 
in \cite{ mom_reach_2, mom_reach_1} to construct reachable sets, and in \cite{mom_motion_1,mom_motion_2, mom_motion_3} for risk bounded motion planning problems (For more examples see \cite{mom_course}). 

In this paper, we address the moment propagation problem through nonlinear stochastic dynamical systems. To compute the moments of uncertain states over the planning horizon, one needs to propagate the moments of the initial uncertain states through the uncertain dynamical system. Several approaches have been proposed to address the moment propagation problem for nonlinear uncertain dynamical systems. However, all the existing methods only provide approximate descriptions of the moments and are mainly limited to particular class of uncertainties. 

For example, linear uncertainty propagation approaches employ linearized models about the current mean and covariance of the uncertain states. Such linear models can be used to propagate the first and second order moments in the presence of additive Gaussian noises as shown in the extended Kalman filter \cite{Robot_Book1, Robot_Book2}. Sampling based approaches, like Monte Carlo simulation, use random samples to approximate the moments. In these approaches, uncertainties are replaced by a large number of particles and, at each time step, moments  are approximated in terms of the propagated particles \cite{Robot_Book1, Robot_Book2}. Due to a large number of samples, these approaches can be computationally intractable. 


Uncertainty propagation based on the unscented transformation uses deterministic samples to approximate and propagate the mean and covariance \cite{mom_UKF, Robot_Book1, UKF_1, Robot_Book2}. In this approach, uncertainties are replaced by a small number of deterministic samples called sigma points. Then, mean and covariance are approximated in terms of the propagated sigma points as shown in the Unscented Kalman filter. Approximated mean and covariance are accurate to the 3rd order of Taylor expansion of nonlinearities in the presence of Gaussian  uncertainties \cite{mom_UKF2, UKF_1}.

In \cite{mom_con_1,mom_opt_1}, we address the uncertainty propagation problem of  polynomial dynamical systems subjected to arbitrary probabilistic uncertainties. In this approach, at each time step, we describe the moments of uncertain states in terms of the moments of the initial uncertain states and process noises using the polynomial moments and polynomial functions obtained by recursion of the dynamical system. This approach is limited to short planning horizons since the order and dimension of the polynomial functions increase as the planning horizon increases. In \cite{mom_Carl}, a recursive approach is provided to approximate the moments of discrete-time stochastic polynomial systems. This approach first transforms the polynomial system into an infinite-dimensional linear system using the Carleman linearization technique. Then, it uses a truncated linear system and polynomial moments to obtain a dynamical system that describes the approximate time evolution of the moments of uncertain states.

Polynomial chaos based uncertainty propagation methods represent the response of stochastic differential systems with an infinite series of orthogonal polynomials with appropriate coefficients \cite{gPc_1,gPc_2, gPc_3, gPc_4}. Such polynomials are functions of uncertain parameters of the stochastic system. Coefficients of the truncated series are used to approximate the mean and covariance of uncertain states.
Numerical issues like the convergence of the series may arise in the presence of non-Gaussian uncertainties and inaccurate truncation orders \cite{gPc_Non_1,gPc_Non_2}. Also, computing coefficients of the polynomial series could be challenging for a large number of uncertainties and highly nonlinear systems and often requires Taylor series approximation of nonlinearities of the dynamical system \cite{gPc_Large_1,gPc_Large_2}.

Moment closure based uncertainty propagation methods look for an approximate closed dynamical system to describe the time evolution of the moments of continuous-time stochastic systems \cite{close_1,close_2}. 
To this end, polynomial moments are used to obtain an infinite dimensional dynamical system in terms of the moments. Then, to obtain a truncated finite system of the moments, higher order moments in the infinite dimensional system are approximated in terms of the lower order moments. 


In this paper, we address discrete-time stochastic nonlinear dynamical systems subjected to arbitrary probabilistic uncertainties. We assume that sources of nonlinearities are mixtures of standard and trigonometric polynomials. Such nonlinearities arise in an important class of autonomous systems including underwater, ground, and aerial vehicles, robotic  arms  and  walking  robots where motion dynamics are described in terms of translational and rotational motions. To be able to address the uncertainty propagation problem and deal with trigonometric and polynomial nonlinearities, we introduce trigonometric and mixed-trigonometric-polynomial moments. These moments allow us to obtain an exact description of the moments of uncertain states for such nonlinear autonomous and robotic systems over the planning horizon. 

Existing methods to address the uncertainty propagation problem of autonomous and robotic systems are all approximate and mainly address Gaussian uncertainties. For example, \cite{robot_Linear} uses linearized models and Gaussian uncertainties to address uncertainty propagation in motion planning of robotic systems under uncertainty, \cite{robot_Sample} uses Monte Carlo based uncertainty propagation for motion planning of NASA's Mars rovers, \cite{robot_Unsc} uses unscented transformation based uncertainty propagation for simultaneous localization and mapping of vehicles, \cite{robot_Unsc2} uses a sample based approach for uncertainty propagation in planetary entry vehicles,  \cite{robot_Chaos} uses polynomial chaos and Gaussian uncertainties for uncertainty propagation in chance constrained motion planning, \cite{close_2} uses moment closure based uncertainty propagation for polynomial and trigonometric stochastic systems in the presence of Gaussian processes, \cite{robot_Linear2} uses linearized models to propagate and represent uncertainties in the form of probabilistic flow tubes, and \cite{robot_Deep} uses deep learning for uncertainty propagation through nonlinear systems. 

In this paper, we leverage trigonometric and mixed-trigonometric-polynomial moments to provide a general framework for uncertainty propagation of stochastic nonlinear autonomous systems. To this end, we will provide two approaches including direct and recursive formulations. In the direct method, we describe the moments of uncertain states in terms of the moments of the initial uncertain states and process noises. In the recursive method, we obtain deterministic linear moment-state dynamical systems that govern the exact time-evolution of the moments of the uncertain states. Hence, to compute the moments, one can recursively propagate the initial moments over the planning horizon in \textit{real-time}. 

The outline of the paper is as follows: in Section \ref{sec_notation}, the notation adopted in the
paper and definitions on polynomials and moments are presented. In Section \ref{sec_formulation}, we provide the problem formulation of moment based nonlinear uncertainty propagation. Section \ref{sec_preliminary} provides preliminary results on nonlinear uncertainty propagation. In Section \ref{sec_mom}, trigonometric and mixed-trigonometric-polynomial moments are introduced. In Section \ref{sec_UTrans}, we use trigonometric and mixed-trigonometric-polynomial moments to compute the moments of nonlinear transformations of uncertainties. We also compare the provided method against existing approaches introduced in Section \ref{sec_preliminary}. In Section \ref{sec_propagation}, we develop direct and recursive approaches to address the uncertainty propagation problem of nonlinear stochastic dynamical systems. 
In Section \ref{sec_results}, we present experimental results on uncertainty propagation for autonomous and  robotic  systems  including underwater, ground, and  aerial vehicles,  robotic  arms  and  walking  robots. Finally, concluding remarks and future work are given in Section \ref{sec_con}.
\section{Notation and Definitions} \label{sec_notation}

\textbf{Standard Polynomials:} Let {\small $\mathbb{R}[x]$} be the set of real polynomials in the variables {\small $\mathbf{x} \in \mathbb{R}^n$}. Given {\small $ P\in\mathbb{R}[x]$}, we represent {\small $P$} as {\small $\sum_{\alpha\in\mathbb{N}^n} p_\alpha x^\alpha$} using the standard monomial basis $x^\alpha = \Pi_{i=1}^n x_i^{\alpha_i}$ where $\alpha=(\alpha_1,...,\alpha_n) \in \mathbb{N}^n$, and {\small $\mathbf{p}=\{p_\alpha\}_{\alpha\in\mathbb{N}^n}$} denotes the coefficients.\\ 

\textbf{Polynomial Moments:} Let $(\Omega,\Sigma_x, \mu_x)$ be a probability space with $\Omega$ denoting the sample space, 
$\Sigma_x$ denoting the $\sigma$-algebra of $\Omega$, and $\mu_x: \Sigma_x \rightarrow [0 \ 1]$ denoting a probability measure on $\Sigma_x$.
Let  $\mathbf{x} \in \mathbb{R}^n$ be a random vector defined on the probability space $\Omega$. Given $(\alpha_1,...,\alpha_n) \in \mathbb{N}^n$ where $\alpha=\sum_{i=1}^{n}\alpha_i$, moment of order $\alpha$ of probability measure $\mu_x$ is defined as $m_{x^{\alpha}}=\mathbb{E}[x^{\alpha}]=\int x^{\alpha} d\mu_x = \int x_1^{\alpha_1}x_2^{\alpha_2}... x_n^{\alpha_n} d\mu_x$. 
As a result, the sequence of all moments of order $\alpha$ is defined as expected values of all monomials of order $\alpha$ sorted according to graded reverse lexicographic order (grevlex). For example, sequence of the moments of order $\alpha=2$ for $n=3$ is defined as $[m_{x_1^{2}}, m_{x_1x_2}, m_{x_1x_3}, m_{x_2^{2}}, m_{x_2x_3}, m_{x_3^{2}}]=$
\begin{small}
 $\left[\mathbb{E}[x_1^{2}], \mathbb{E}[x_1x_2],
\mathbb{E}[x_1x_3],
\mathbb{E}[x_2^2] ,
\mathbb{E}[x_2x_3],
\mathbb{E}[x_3^2] \right]$\end{small}.

Moments of a probability measure can be directly
computed using it's characteristic function. The characteristic function is the Fourier transform of the probability density function of a random variable. For any probability density function, the characteristic function  always exists \cite{Char_1} and is defined as:
\begin{align} \label{CF}
    \Phi_{\mathbf{x}}(t) = \mathbb{E}[e^{it^T\mathbf{x}}]
\end{align}
Moment of order $\alpha$ can be computed by applying $n$ partial derivatives as follows \cite{Char_2}:
\begin{equation} \label{poly_mom_1}
    m_{x^{\alpha}}= i^{-(\alpha_1+...+\alpha_n)} \frac{\partial^{\alpha_1+...+\alpha_n}}{\partial t_1^{\alpha_1}...\partial t_n^{\alpha_n}}  \Phi_{\mathbf{x}}(t_1,...,t_n)|_{t_1=0,...,t_n=0}
\end{equation}

\textbf{Trigonometric Polynomials:} Trigonometric polynomials of order $n$ are defined as $P(x)=\sum_{k=0}^n p_{c_k} cos(kx)+p_{s_k}sin(kx) $ where $\{p_{c_k}, p_{s_k}\}_{k=0}^{n}$ are the coefficients\cite{Tri_1}.\\

\textbf{Mixed Trigonometric Polynomials:} Mixed trigonometric polynomials are a mixture of standard and trigonometric polynomials and are defined as  $P(x)=\sum_{k=0}^n p_k x^{a_k}cos^{b_k}(x)sin^{c_k}(x)$ where $a_k, b_k, c_k \in \mathbb{N}$ and $\{p_k\}_{k=0}^{n}$ are the coefficients \cite{Mixed_Tri_1, Mixed_Tri_2}.

\section{Problem Formulation} \label{sec_formulation}
Consider the discrete-time continuous-state stochastic nonlinear dynamical system of the form:
\begin{equation} \label{sys}
\mathbf{x}(k+1)=f(\mathbf{x}(k),\mathbf{u}(k),\mathbf{\omega}(k))    
\end{equation}
where $\mathbf{x}(k) \in \mathbb{R}^{n_x}$ is the state vector, $\mathbf{u}(k) \in \mathbb{R}^{n_u}$ is the input vector, and $\mathbf{\omega}(k) \in \mathbb{R}^{n_{\omega}}$ is the probabilistic process noise vector at time step $k$. Sources of uncertainties are probabilistic initial states $\mathbf{x}(0)$ and process noise $\mathbf{\omega}(k)$ with given arbitrary probability distributions. In this paper, we aim at solving the moment based nonlinear uncertainty propagation problem defined as follows:

\textbf{Uncertainty Propagation Problem:} Given the stochastic nonlinear system in \eqref{sys} and nominal control inputs over the planning horizon, i.e., $\mathbf{u}^*(k),k=0,...,N-1$, we want to propagate the finite sequence of the moments of the initial probability distributions of the states, i.e., $m_{x^{\alpha}}(0)=\mathbb{E}[x^{\alpha}(0)], \alpha=1,...,d$, through the nonlinear stochastic dynamical system and obtain the moments of the probability distributions of the states over the planning horizon, i.e.,
$m_{x^{\alpha}}(k)=\mathbb{E}[x^{\alpha}(k)], \alpha=1,...d, \ k=1,...,N$.

In this paper, we assume that all the nonlinear transformations take the standard, trigonometric, or mixed-trigonometric polynomial forms defined in Section \ref{sec_notation}. 
Such nonlinear transformations arise in many different systems. This includes an important class of autonomous and robotic systems such as underwater, ground, and aerial vehicles, robotic arms and walking robots. In such systems, function $f$ of dynamics \eqref{sys} captures the translational and rotational motions of the system. Hence,  $f$ includes linear and trigonometric terms, i.e., rotation matrix, to describe the translational and rotational motions, respectively (e.g., see Section \ref{sec_results}).



\section{Preliminary Results on Nonlinear Uncertainty Propagation} \label{sec_preliminary}
In this section, we review linear, Monte Carlo simulation, and unscented transformation based uncertainty propagation methods that are widely used in estimation, prediction, planning, and control problems. To illustrate the performance of our proposed method, we will benchmark our method against these existing approaches.
\subsection{Linear Uncertainty Propagation:}

This approach uses linearized models about the current mean and covariance of the uncertain states, e.g., extended Kalman filter \cite{Robot_Book1, Robot_Book2}.
More precisely, given the nonlinear model in \eqref{sys} with zero mean additive Gaussian noise $\mathbf{\omega}(k)$ with covariance matrix $Q(k)$, the update rule to propagate the mean and covariance of states $\mathbf{x}(k)$ are as follows:
\begin{align} \label{EKF}
\begin{split}
\hat{\mathbf{x}}(k+1) &=f(\hat{\mathbf{x}}(k),\mathbf{u}(k))\\
P(k+1) &=A(k)P(k)A^T(k) + Q(k)
\end{split}
\end{align} 
where, $\hat{\mathbf{x}}$ is the mean vector and $P$ is the covariance matrix of the uncertain states. Also, $A$ is the Jacobian matrix defined as $A(k)=\frac{\partial f}{\partial x}|_{\hat{\mathbf{x}}(k),\mathbf{u}(k)}$. 

\subsection{Monte Carlo Simulation Based Uncertainty Propagation:}
Monte Carlo simulation uses random samples to approximate the moments of the uncertain states at each time step $k$. More precisely, given the nonlinear model in \eqref{sys}, let \begin{small} $\{\mathbf{x}^{(i)}(0)\}_{i=1}^{N_s}$ \end{small}  and \begin{small}$\{\mathbf{\omega}^{(i)}(k)\}_{i=1}^{N_s}$\end{small}  be 
the samples obtained from the distributions of the initial uncertain states $\mathbf{x}(0)$ and the process noise $\mathbf{\omega}(k)$, respectively.
Then, the propagated samples of the uncertain states reads as:
\begin{equation}
\mathbf{x}^{(i)}(k+1)=f(\mathbf{x}^{(i)}(k),\mathbf{\omega}^{(i)}(k),\mathbf{u}(k)), \ i=1,...,N_s    
\end{equation}
Moment of order $\alpha$ at time step $k+1$ can be approximated in terms of the propagated samples as follows:
\begin{equation} \label{mom_sample}
\mathbb{E}[x^{\alpha}(k+1)] \approx \frac{1}{N_s}\sum_{i=1}^{N_s} \left( x^{(i)}(k+1) \right)^{\alpha}
\end{equation}
Because of the law of large numbers, as the number of samples $N_s$ increases, the approximated moment in \eqref{mom_sample} converges to the true value of the moment of the uncertain states.

\subsection{Unscented Transformation Based Uncertainty Propagation:}
This approach, unlike Monte Carlo simulation, uses deterministic samples to represent the uncertainty and propagate the mean and covariance of the uncertain states, e.g., Unscented Kalman filter \cite{UKF_1, mom_UKF, mom_UKF2}. More precisely, given the nonlinear model in \eqref{sys}, uncertainty vector $\mathbf{z}(k)=[\mathbf{x}^T(k) \ \mathbf{\omega}^T(k)]^T\in \mathbb{R}^{n_x+n_{\omega}}$ is approximated by $2(n_x+n_{\omega})+1$ deterministic samples called sigma points as follows:
\begin{align} \label{UKF1}
 \mathbf{z}^{(i)}(k) &= \hat{\mathbf{z}}(k), \hfill  \ \ \ \ \ \ \ \ \ \ \ \ \ \ \ \ \ \ \ \ \  i=0 \\
 \mathbf{z}^{(i)}(k) &= \hat{\mathbf{z}}(k) + \left( \sqrt{cP(k)} \right)_i, \ \ i=1,...,n_x+n_{\omega} \nonumber \\
\mathbf{z}^{(i+n_x+n_{\omega})}(k) &= \hat{\mathbf{z}}(k) - \left( \sqrt{cP(k)} \right)_i, \ \ i=1,...,n_x+n_{\omega} \nonumber
\end{align} 
where, $\hat{\mathbf{z}}$ and $P$ are the mean vector and covariance matrix of vector $\mathbf{z}$, $c=n_x+n_{\omega}+\kappa$ and $ \kappa \in \mathbb{R}$ are scaling factors, and \begin{small} $\left( \sqrt{cP(k)} \right)_i$ \end{small} is the $i$-th column of the matirx square root of $cP(k)$. Then, the  update  rule  to  propagate  the  mean  and covariance of the uncertainty vector $\mathbf{z}$ in terms of the propagated sigma points reads as:
\begin{small}
\begin{align} \label{UKF2}
 \hat{\mathbf{z}}(k+1) = & \sum_{i=0}^{2(n_x+n_{\omega})} W_i \mathbf{z}^{(i)}(k+1) \\
P(k+1) = & 
\sum_{i=0}^{2(n_x+n_{\omega})} W_i \left( \mathbf{z}^{(i)}(k+1) - \hat{\mathbf{z}}(k+1) \right) \left( \mathbf{z}^{(i)}(k+1) - \hat{\mathbf{z}}(k+1) \right)^T  \nonumber
\end{align} 
\end{small}

\noindent where the weights are defined as \begin{small} $W_0=\frac{\kappa}{n_x+n_{\omega}+\kappa},\  W_i=\frac{1}{2(n_x+n_{\omega}+\kappa)}, \ i=1,...,2(n_x+n_{\omega})$\end{small}. The approximated mean and covariance are accurate to the 3rd order of Taylor expansion of nonlinearities in the presence of Gaussian  uncertainties \cite{UKF_1, mom_UKF, mom_UKF2}.

\section{Trigonometric and Mixed-Trigonometric-Polynomial Moments} \label{sec_mom}
In this section, we 
extend the notion of polynomial moments introduced in Section \ref{sec_notation} to define trigonometric and mixed-trigonometric-polynomial moments. We will use such moments to deal  with trigonometric  and  polynomial nonlinearities and address the nonlinear uncertainty propagation problem. We will show how to compute trigonometric and mixed-trigonometric-polynomial moments using the characteristic function of uncertainties.

\subsection{Trigonometric Moments}
Given $(\alpha_1,\alpha_2) \in \mathbb{N}^2$ where $\alpha=\sum_{i=1}^{2}\alpha_i$, we define the trigonometric moment of order $\alpha$ for random variable $\theta$ as follows:
\begin{equation} \label{Tri_mom_1}
m_{c^{\alpha_1}_{\theta}s^{\alpha_2}_{\theta}}=\mathbb{E}[{cos}^{\alpha_1}(\theta){sin}^{\alpha_2}(\theta)] \  \end{equation}

Trigonometric moments of the form \eqref{Tri_mom_1} can be computed in terms of the characteristic function of the random variable $\theta$. The following Lemmas allow us to compute trigonometric moments of a random variable in terms of the linear sum of its characteristic function.\\

\textbf{Lemma 1:} Let $\theta$ be a random variable with characteristic function $\Phi_{\theta}(t)$. Given $\alpha \in \mathbb{N}, $ trigonometric moments of order $\alpha$ of the forms $m_{c^{\alpha}_{\theta}}= \mathbb{E}[{cos}^{\alpha}(\theta)]$ and $  m_{s^{\alpha}_{\theta}}=\mathbb{E}[{sin}^{\alpha}(\theta)] $ reads as:
\begin{align} \label{Tri_mom_2}
m_{c_{\theta}^{\alpha}}=\mathbb{E}[{cos}^{\alpha}(\theta)] =\frac{1}{2^{\alpha}} \sum_{k=0}^{\alpha} \binom{\alpha}{k}\Phi_{\theta}(2k-\alpha)
\end{align}
\begin{align} \label{Tri_mom_3}
m_{s_{\theta}^{\alpha}}=\mathbb{E}[{sin}^{\alpha}(\theta)]&=\frac{(-i)^{\alpha}}{2^{\alpha}} \sum_{k=0}^{\alpha} \binom{\alpha}{k}(-1)^{\alpha-k}\Phi_{\theta}(2k-\alpha)
\end{align}
where, $\binom{\alpha}{k}=\frac{\alpha !}{k!(\alpha-k)!}$. \\

\textit{Proof}: By applying Euler's Identity and Binomial formula, we obtain:
\begin{small} $$\mathbb{E} [ {cos}^{\alpha}(\theta)] =\mathbb{E} \left[\left(\frac{e^{i\theta}+e^{-i\theta}}{2}\right)^{\alpha} \right]= \frac{1}{2^{\alpha}} \sum_{k=0}^{\alpha} \binom{\alpha}{k} \mathbb{E} \left[ e^{i(2k-\alpha)\theta} \right]
$$ \end{small}
Then, we use the definition of characteristic function to replace $\mathbb{E}[e^{i(2k-\alpha)\theta}]$ with $\Phi_{\theta}(t=2k-\alpha)$ and obtain the trigonometric moment in \eqref{Tri_mom_2}. Similarly, we can obtain the trigonometric moment in \eqref{Tri_mom_3}.
 \hfill  $\square$ \\

\textbf{Lemma 2:} Let $\theta$ be a random variable with characteristic function $\Phi_{\theta}(t)$. Given $(\alpha_1,\alpha_2) \in \mathbb{N}^2$ where $\alpha=\sum_{i=1}^2 \alpha_i$, trigonometric moment of order $\alpha$ of the form $m_{c^{\alpha_1}_{\theta}s^{\alpha_2}_{\theta}}=\mathbb{E}\left[{cos}^{\alpha_1}(\theta){sin}^{\alpha_2}(\theta) \right]$ reads as:
\begin{equation} \label{Tri_mom_4}
 m_{c^{\alpha_1}_{\theta}s^{\alpha_2}_{\theta}}=\mathbb{E}\left[{cos}^{\alpha_1}(\theta){sin}^{\alpha_2}(\theta) \right]= \frac{(-i)^{\alpha_2}}{2^{\alpha_1+\alpha_2}} \sum_{(k_1,k_2)=(0,0)}^{(\alpha_1,\alpha_2)}  \binom{\alpha_1}{k_1}\binom{\alpha_2}{k_2}(-1)^{\alpha_2-k_2}\Phi_{\theta}\left(2(k_1+k_2)-\alpha_1-\alpha_2\right)
\end{equation}


\textit{Proof}: Similar to Lemma 1, we obtain the trigonometric moment in \eqref{Tri_mom_4} by applying Euler's Identity and Binomial formula. \hfill  $\square$ \\

\textbf{Illustrative Example 1:} Let $\theta$ be a random variable with uniform probability distribution over $[0,0.5]$.  Characteristic function of uniform probability distribution over $[l,u]$ reads as $\Phi_{\theta}(t)=\frac{e^{itu}-e^{itl}}{it(u-l)}$.
Using \eqref{Tri_mom_2} and \eqref{Tri_mom_3}, we obtain the first order trigonometric moments of $\theta$ as:
\begin{align} 
\begin{split}
 m_{c_{\theta}}=\mathbb{E}[cos(\theta)]  =\frac{1}{2}\Phi_{\theta}(-1)+  \frac{1}{2}\Phi_{\theta}(1)=0.9589, \ \ \ \ 
   m_{s_{\theta}}=\mathbb{E}[sin(\theta)]=\frac{i}{2}\Phi_{\theta}(-1)+  \frac{-i}{2}\Phi_{\theta}(1)=0.2448 \notag
\end{split}
\end{align}
Using \eqref{Tri_mom_4}, we obtain the second order trigonometric moment of the form $m_{c_{\theta}s_{\theta}}$ as:
\begin{equation}
    m_{c_{\theta}s_{\theta}}=\mathbb{E}[cos(\theta)sin(\theta)]=
   \frac{i}{4} \left( \Phi_{\theta}(-2)+
   \Phi_{\theta}(0)
   -\Phi_{\theta}(0)
   -\Phi_{\theta}(2) \right) = 0.229 \nonumber
\end{equation}
Similarly, we obtain the sequence of the all third order trigonometric moments of $\theta$ as follows: 
\begin{align} 
\begin{split}
&\left[ m_{c^3_{\theta}}, m_{c^2_{\theta}s_{\theta}},m_{c_{\theta}s^2_{\theta}},m_{s^3_{\theta}} \right]
 =\left[0.8854,0.2161,0.0735,0.0287\right] \notag
\end{split}
\end{align} 
One can verify the results using the Monte Carlo simulation.
\subsection{Mixed-Trigonometric-Polynomial Moments}
Given $(\alpha_1,\alpha_2,\alpha_3) \in \mathbb{N}^3$ where $\alpha=\sum_{i=1}^{3}\alpha_i$, we define the mixed-trigonometric-polynomial moment of order $\alpha$ for random variable $\theta$ as follows:
\begin{equation} \label{Mixed_Tri_mom_1}
m_{{\theta^{\alpha_1} c^{\alpha_2}_{\theta}s^{\alpha_3}_{\theta}}}=\mathbb{E}[\theta^{\alpha_1} {cos}^{\alpha_2}(\theta){sin}^{\alpha_3}(\theta)] \  \end{equation}

Mixed-trigonometric-polynomial moments of the form \eqref{Mixed_Tri_mom_1} can be computed in terms of the characteristic function of the random variable $\theta$. The following Lemmas allow us to compute mixed-trigonometric-polynomial moments of a random variable in terms of the linear sum of derivatives of its characteristic function. \\

\textbf{Lemma 3:} Let $\theta$ be a random variable with characteristic function $\Phi_{\theta}(t)$. Given $(\alpha_1,\alpha_2) \in \mathbb{N}^2$ where $\sum_{i=1}^2 \alpha_i = \alpha$, mixed-trigonometric-polynomial moments of order $\alpha$ of the forms \begin{small}$m_{\theta^{\alpha_1} c^{\alpha_2}_{\theta}}= \mathbb{E}\left[\theta^{\alpha_1}{cos}^{\alpha_2}(\theta)\right]$ and $ m_{\theta^{\alpha_1} s^{\alpha_2}_{\theta}}=\mathbb{E}\left[\theta^{\alpha_1}{sin}^{\alpha_2}(\theta)\right]$ \end{small} reads as:
\begin{align} \label{Mixed_Tri_mom_2}
m_{\theta^{\alpha_1}c_{\theta^{\alpha_2}}}& =\mathbb{E}[\theta^{\alpha_1}{cos}^{\alpha_2}(\theta)] = 
 \frac{1}{i^{\alpha_1} 2^{\alpha_2}} \sum_{k=0}^{\alpha_2} \binom{\alpha_2}{k}\frac{d^{\alpha_1} }{d t^{\alpha_1}} \Phi_{\theta}(t)|_{t=2k-\alpha_2}
\end{align}
\begin{align} \label{Mixed_Tri_mom_3}
m_{\theta^{\alpha_1}s_{\theta^{\alpha_2}}}& =\mathbb{E}  \left[\theta^{\alpha_1}{sin}^{\alpha_2}(\theta)  \right]= 
 \frac{1}{i^{\alpha_1+\alpha_2}2^{\alpha_2}} \sum_{k=0}^{\alpha_2} \binom{\alpha_2}{k}(-1)^{\alpha_2-k}\frac{d^{\alpha_1} }{d t^{\alpha_1}} \Phi_{\theta}(t)|_{t=2k-\alpha_2}
\end{align}

\textit{Proof}: By applying Euler's Identity and Binomial formula, we obtain:
\begin{equation}
\mathbb{E}\left[ \theta^{\alpha_1} {cos}^{\alpha_2}(\theta) \right] = \mathbb{E}  \left[ \theta^{\alpha_1}  \left(\frac{e^{i\theta}+e^{-i\theta}}{2}\right)^{\alpha_2}  \right] 
= \mathbb{E}  \left[ \theta^{\alpha_1} \frac{1}{2^{\alpha_2}} \sum_{k=0}^{\alpha_2} \binom{\alpha_2}{k}e^{i(2k-\alpha_2)\theta} \right] \nonumber    
\end{equation}

Then, we replace \begin{small}$\mathbb{E}\left[\theta^{\alpha_1}e^{it\theta} \right]|_{t=2k-\alpha_2} $\end{small} with $\mathbb{E}\left[ \frac{1}{i^{\alpha_1}}  \frac{d^{\alpha_1}}{dt^{\alpha_1}} e^{it\theta} \right] |_{t=2k-\alpha_2}$. As shown in \cite{Mixed_Tri_3}, we can interchange the derivative and expectation. Hence,
\begin{small}
$ \mathbb{E}\left[ \frac{1}{i^{\alpha_1}}  \frac{d^{\alpha_1}}{dt^{\alpha_1}} e^{it\theta} \right]|_{t=2k-\alpha_2}$ = $  \frac{1}{i^{\alpha_1}}  \frac{d^{\alpha_1}}{dt^{\alpha_1}} \mathbb{E}\left[e^{it\theta} \right]|_{t=2k-\alpha_2} $.
\end{small}
Then, we use the definition of characteristic function to replace $\mathbb{E}[e^{it\theta}]$ with $\Phi_{\theta}(t)$ and obtain the mixed-trigonometric-polynomial moment in \eqref{Mixed_Tri_mom_2}. Similarly, we can obtain the mixed-trigonometric-polynomial moment in \eqref{Mixed_Tri_mom_3}.
 \hfill  $\square$ \\
 


\textbf{Lemma 4:} Let $\theta$ be a random variable with characteristic function $\Phi_{\theta}(t)$. Given $(\alpha_1,\alpha_2,\alpha_3) \in \mathbb{N}^3$ where $\sum_{i=1}^3 \alpha_i = \alpha$, mixed-trigonometric-polynomial moment of order $\alpha$ of the form $m_{\theta^{\alpha_1} c^{\alpha_2}_{\theta}s^{\alpha_3}_{\theta}}= \mathbb{E}\left[\theta^{\alpha_1}{cos}^{\alpha_2}(\theta){sin}^{\alpha_3}(\theta) \right]$ reads as:

\begin{small}
\begin{align} \label{Mixed_Tri_mom_4}
 m_{\theta^{\alpha_1}c^{\alpha_2}_{\theta}s^{\alpha_3}_{\theta}}=\mathbb{E}\left[\theta^{\alpha_1}{cos}^{\alpha_2}(\theta){sin}^{\alpha_3}(\theta) \right]= \frac{1}{i^{\alpha_1+\alpha_3}2^{\alpha_2+\alpha_3}} \times 
  \sum_{(k_1,k_2)=(0,0)}^{(\alpha_2,\alpha_3)}  \binom{\alpha_2}{k_1}\binom{\alpha_3}{k_2}(-1)^{\alpha_3-k_2}\frac{d^{\alpha_1}}{dt^{\alpha_1}}\Phi_{\theta}(t)|_{t=2(k_1+k_2)-\alpha_2-\alpha_3}
\end{align}
\end{small}

\textit{Proof}: By applying Euler's Identity and Binomial formula to $cos^{\alpha_2}(\theta)$ and $sin^{\alpha_3}(\theta)$, we obtain:
 
\begin{small}
\begin{align} 
\begin{split}
 \mathbb{E}  \left[ \theta^{\alpha_1}{cos}^{\alpha_2}(\theta){sin}^{\alpha_3}(\theta) \right] & = 
   \mathbb{E}  \left[ \frac{(-i)^{\alpha_3}}{2^{\alpha_2+\alpha_3}} \sum_{(k_1,k_2)=(0,0)}^{(\alpha_2,\alpha_3)}  \binom{\alpha_2}{k_1}\binom{\alpha_3}{k_2}(-1)^{\alpha_3-k_2}\theta^{\alpha_1}e^{it\theta}  \right]\\
   & = \mathbb{E}  \left[ \frac{(-i)^{\alpha_3}}{i^{\alpha_1}2^{\alpha_2+\alpha_3}} \sum_{(k_1,k_2)=(0,0)}^{(\alpha_2,\alpha_3)}  \binom{\alpha_2}{k_1}\binom{\alpha_3}{k_2}(-1)^{\alpha_3-k_2}\frac{d^{\alpha_1}}{dt^{\alpha_1}}e^{it\theta} \right] \notag
\end{split}
\end{align}
\end{small} 

\noindent for $t=2(k_1+k_2)-\alpha_2-\alpha_3$. Similar to the proof of Lemma 3, we interchange the derivative and expectation and use the definition of characteristic function to obtain the mixed-trigonometric-polynomial moment in \eqref{Mixed_Tri_mom_4}.
 \hfill  $\square$\\

\textit{Remark 1}: Note that polynomial and trigonometric moments are particular cases of mixed-trigonometric-polynomial moments.
Hence, in the remainder of this paper, we use mixed-trigonometric-polynomial moments to refer to polynomial and trigonometric moments as well.\\

\textbf{Illustrative Example 2:} Consider the random variable in illustrative example 1.
Using \eqref{Mixed_Tri_mom_2} and \eqref{Mixed_Tri_mom_3}, we obtain second order mixed-trigonometric-polynomial moments of the form 
\begin{small} $\mathbb{E}[\theta cos(\theta)]$\end{small} and \begin{small}$\mathbb{E}[\theta sin(\theta)]$\end{small}
as:
\begin{align} 
\begin{split}
m_{\theta c_{\theta}}=\mathbb{E}[\theta cos(\theta)]=\frac{1}{2i} \frac{d}{dt} \Phi_{\theta}(t)|_{t=-1} +\frac{1}{2i}\frac{d}{dt}\Phi_{\theta}(t)|_{t=1} = 0.234 \\
m_{\theta s_{\theta}}=\mathbb{E}[\theta sin(\theta)]=\frac{1}{2} \frac{d}{dt} \Phi_{\theta}(t)|_{t=-1} +\frac{1}{-2}\frac{d}{dt}\Phi_{\theta}(t)|_{t=1}= 0.081
\notag
\end{split}
\end{align}
Using \eqref{Mixed_Tri_mom_4}, we obtain forth order mixed-trigonometric-polynomial moment of the form \begin{small} $\mathbb{E}\left[\theta^2 cos(\theta) sin(\theta)\right]$\end{small} as:
\begin{equation}
m_{\theta^2 c_{\theta} s_{\theta}}= \mathbb{E}[\theta^2 cos(\theta) sin(\theta)]= \frac{1}{4i} \left(  \frac{d^2}{dt^2} \Phi_{\theta}( -t ) 
- \frac{d^2}{dt^2} \Phi_{\theta}( t )
\right)|_{t=2} = 0.027 \nonumber
\end{equation}

One can verify the results using the Monte Carlo simulation.


\section{Nonlinear Uncertainty Transformation} \label{sec_UTrans}

In this section, we use mixed-trigonometric-polynomial moments defined in Section \ref{sec_mom} to compute the moments of nonlinear transformations of uncertainties. More precisely, 
let $\mathbf{x} \in \mathbb{R}^n$ be a random vector with given probability measure $\mu_x$. Let $z=f(\mathbf{x})$ be a nonlinear transformation of the random vector $\mathbf{x}$ with respect to function $f: \mathbb{R}^n \rightarrow \mathbb{R}$. In this section, we aim at finding the moments of the new random variable $z \in \mathbb{R}$, i.e., $m_{z^{\alpha}}=\mathbb{E}[z^{\alpha}], \ \forall \alpha \in \mathbb{N}$. We can describe the moments of the random variable $z$ in terms of the moments of the random vector $\mathbf{x}$ by defining the pusforward measure of function $f$ as follows: \\

\textbf{Pushforward Measure:} Pushforward measure on $\mathbb{R}$ of probability measure $\mu_x$ with respect to function $f: \mathbb{R}^n \rightarrow \mathbb{R}$ denoted by $ \# \mu_x $ is defined as follows \cite{Push_1,Push_2}:
\begin{equation}
\# \mu_x(S) = \mu_x (f^{-1}(S)) \ \ \ \forall S \in \mathcal{B}(\mathbb{R})   
\end{equation}
where, $\mathcal{B}(\mathbb{R})$ is the $\sigma$-algebra of $\mathbb{R}$.
In fact, the pushforwrad measure $ \# \mu_x $ represents the transformation of the probability measure $\mu_x$ under the given function $f$, i.e., probability measure of random variable $z$. The moment of order $\alpha$ of the random variable $z$ can be described in terms of the pushforwrad measure $ \# \mu_x $ as $m_{z^{\alpha}}=\int z^{\alpha} d (\# \mu_x)$ \cite{mom_risk_1,Push_2,Push_1}. Hence, we can describe the moments of $z$ in terms of the moments of $\mathbf{x}$ as follows:
\begin{equation} \label{pushforward}
m_{z^{\alpha}}=\int z^{\alpha} d (\# \mu_x)=\int f^{\alpha}(\mathbf{x}) d\mu_x
\end{equation}
We assume that nonlinear function $f$ takes the form of standard, trigonometric, or mixed trigonometric polynomials. Hence, according to \eqref{pushforward}, we can compute the moments of $z$ in terms of the mixed-trigonometric-polynomial moments of $\mathbf{x}$.\\


\textbf{Illustrative Example 3} - \textit{Polar to Cartesian Coordinate Transformation:} Suppose a LiDAR sensor of a mobile robot detects an obstacle in the environment. The sensor returns noisy polar information including range $r^*+\omega_r$ and bearing $\theta^*+\omega_{\theta}$, where $(r^*=1,\theta^*=\pi/2)$ are the actual range and bearing and  $\omega_r$ and $\omega_{\theta}$ are independent measurement noises \cite{UKF_1}. Cartesian coordinates of the obstacle reads as:
\begin{align} \label{Cart}
\begin{split}
x =(r^*+\omega_r)cos(\theta^*+\omega_{\theta}), \ \ 
y =(r^*+\omega_r)sin(\theta^*+\omega_{\theta}) \notag
\end{split}
\end{align} 
We want to find the moments of the location of the obstacle in the Cartesian coordinates, i.e., $m_{x^{\alpha_1}y^{\alpha_2}}=\mathbb{E}[x^{\alpha_1}y^{\alpha_2}], \  \forall (\alpha_1,\alpha_2)\in \mathbb{N}^2$. Cartesian moments read as:
\begin{equation}
\mathbb{E}[x^{\alpha_1}y^{\alpha_2}]=\mathbb{E}\left[ \left( r^*+\omega_{r} \right)^{\alpha_1+\alpha_2}\right] \mathbb{E} \left[cos^{\alpha_1}(\theta^*+\omega_{\theta})sin^{\alpha_2}(\theta^*+\omega_{\theta}) \right] \notag
\end{equation}

Using the trigonometric power and Binomial formula, Cartesian moments read as:
\begin{equation}
    \mathbb{E}[x^{\alpha_1}y^{\alpha_2}]=
  (-1)^{\alpha_1} \left( \sum_{k=0}^{\alpha_1+\alpha_2} \binom{\alpha_1+\alpha_2}{k}  \mathbb{E}\left[\omega_r^{k} \right]  \right)  \mathbb{E} \left[ cos^{\alpha_2} (\omega_{\theta}) sin^{\alpha_1} (\omega_{\theta})     \right] \nonumber
\end{equation}

\noindent where  \begin{small} $\mathbb{E}\left[cos^{\alpha_2}(\omega_{\theta})sin^{\alpha_1}(\omega_{\theta}) \right] $ \end{small} is the trigonometric moment of $\omega_{\theta}$ defined in \eqref{Tri_mom_1} and can be computed in terms of its characteristic function as in \eqref{Tri_mom_4}. Similarly, $\mathbb{E}\left[ \omega_r^k\right]$ is the polynomial moment of $\omega_r$ and can be computed in terms of its characteristic function as in \eqref{poly_mom_1}. We use first ad second order Cartesian moments to obtain the mean and variance of the location of the obstacle. In Table \ref{tab_1}, we show the obtained mean and variance of Cartesian coordinates for the different probability distributions of the measurement noises $\omega_r$ and $\omega_{\theta}$ including normal, uniform, and Beta distributions. 

We compare the proposed moment based approach with i) linear,  ii) unscented transformation, and iii) Monte Carlo simulation based methods. Using our proposed moment formulation, we obtain the true values of the mean and variance verified by the Monte Carlo simulation with $10^8$ samples. Also, as shown in Table \ref{tab_1}, unscented transformation based approach gives accurate results only in the presence of normal distributions with small variances, e.g., case I. But, accuracy decreases in the presence of normal distributions with large variance as in case II and in the presence of non-Gaussian distributions as in case III.\\

\begin{table} 
\renewcommand{\arraystretch}{1.2}
\begin{scriptsize}
    \begin{center}
    \begin{tabular}{| l | l | l | l | l |} \hline
    \multicolumn{5}{|c|}{Case I: $\omega_r \sim \mathcal{N}(0,0.02^2)$, $\omega_{\theta} \sim \mathcal{N}(0,0.2^2)$  }  \\ \hline
    & Linear & Unscented & Moment & Monte Carlo ($N_s= 10^8$) \\ \hline
    $\mathbb{E}$[$x$] & 0 &  0 &  0 & 0 \\ \hline
    $\mathbb{E}$[$y$] & 1 &  0.9802 &  0.9802 & 0.9802 \\ \hline
    $ \mathbb{E}[x^2]-\mathbb{E}[x]^2 $ &  0.04  &  0.0384&   0.0385 &  0.0385  \\ \hline  
    $ \mathbb{E}[y^2]-\mathbb{E}[y]^2 $ &  0.0004 & 0.0012 & 0.0012 & 0.0012  \\ \hline  \hline
    
    \multicolumn{5}{|c|}{Case II: $\omega_r \sim \mathcal{N}(0,0.3^2)$, $\omega_{\theta} \sim \mathcal{N}(0,1)$ } \\ \hline
    & Linear & Unscented & Moment & Monte Carlo ($N_s= 10^8$) \\ \hline
    $\mathbb{E}$[$x$] & 0 &  0 &  0 & 0 \\ \hline
    $\mathbb{E}$[$y$] & 1 &    0.613 &  0.606 &   0.606 \\ \hline
    $ \mathbb{E}[x^2]-\mathbb{E}[x]^2 $ &  1  &  0.324 &  0.471 &  0.471  \\ \hline  
    $ \mathbb{E}[y^2]-\mathbb{E}[y]^2 $ &  0.009 & 0.389 & 0.250 & 0.250 \\ \hline  \hline
    
       \multicolumn{5}{|c|}{Case III: $\omega_r \sim \mathcal{B}eta(3, 0.1)$, $\omega_{\theta} \sim \mathcal{U}niform(-2,2)$ } \\ \hline
    & Linear & Unscented & Moment & Monte Carlo ($N_s= 10^8$) \\ \hline
    $\mathbb{E}$[$x$] & 0 &  0 &  0 & 0 \\ \hline
    $\mathbb{E}$[$y$] &  1.967  &  1.038   & 0.894 &  0.894 \\ \hline
    $ \mathbb{E}[x^2]-\mathbb{E}[x]^2 $ &  5.1627  &    1.0672   &  2.3068 &  2.3068   \\ \hline  
    $ \mathbb{E}[y^2]-\mathbb{E}[y]^2 $ &  0.0076    &   1.7332   & 0.772 &   0.772  \\ \hline 
   
    \end{tabular}
    \end{center}
\end{scriptsize}
\vspace{3 mm}
\caption{Illustrative Example 3 - Cartesian mean and variance of noisy measurements}
\label{tab_1}
\end{table}

\textbf{Illustrative Example 4} - \textit{Non-Additive Noise}:
A process noise is modeled by filtering the external noise $\eta$ through a nonlinear filter as $\omega=0.9\eta^3+\eta$. We can compute the moments of $\omega$ in terms of the moments of $\eta$ as follows:
\begin{equation}
\mathbb{E}[\omega^{\alpha}]=\mathbb{E}\left[\left( 0.9\eta^3+\eta \right)^{\alpha} \right]=
\sum_{k=0}^{\alpha} \binom{\alpha}{k} 0.9^k \mathbb{E}\left[\eta^{2k+\alpha}\right]   \notag 
\end{equation}
In Table \ref{tab_2}, we show the obtained mean and variance of $\omega$ for the different probability distributions of the external noise $\eta$ including normal, uniform, and Beta distributions. 
We also compare the proposed moment based approach with i) linear,  ii) unscented transformation, and iii) Monte Carlo simulation based methods. Unlike the linear and unscented transformation based methods, the proposed moment based approach obtains the true values of the mean and variance verified by the Monte Carlo simulation  with $10^8$ samples. 




\begin{table} 
\renewcommand{\arraystretch}{1.2}
\begin{scriptsize}
    \begin{center}
    \begin{tabular}{| l | l | l | l | l |} \hline
    \multicolumn{5}{|c|}{Case I: $\eta \sim \mathcal{N}(0,0.1)$  }  \\ \hline
    & Linear & Unscented & Moment & Monte Carlo ($N_s= 10^8$) \\ \hline
    $\mathbb{E}$[$x$] & 0 &  0 &  0 & 0 \\ \hline
    $ \mathbb{E}[x^2]-\mathbb{E}{E}[x]^2 $ &  0.10  &  0.161  &  0.166 & 0166
      \\ \hline  \hline
    
    \multicolumn{5}{|c|}{Case II: $\eta \sim \mathcal{N}(0,0.5)$ } \\ \hline
    & Linear & Unscented & Moment & Monte Carlo ($N_s= 10^8$) \\ \hline
    $\mathbb{E}$[$x$] & 0 &  0 &  0 & 0 \\ \hline
    $ \mathbb{E}[x^2]-\mathbb{E}[x]^2 $ &  0.5  &  2.761    & 3.367 &  3.368 \\ \hline  \hline
    
       \multicolumn{5}{|c|}{Case III: $\eta \sim \mathcal{U}niform(-0.5,0.5)$ } \\ \hline
    & Linear & Unscented & Moment & Monte Carlo ($N_s= 10^8$) \\ \hline
    $\mathbb{E}$[$x$] & 0 &  0 &  0 & 0 \\ \hline
    $ \mathbb{E}[x^2]-\mathbb{E}[x]^2 $ &  0.0833 & 0.125 & 0.107 & 0.107 \\ \hline  \hline
    
    \multicolumn{5}{|c|}{Case IV: $\eta \sim \mathcal{B}eta(0.75,0.75)$ } \\ \hline
    & Linear & Unscented & Moment & Monte Carlo ($N_s= 10^8$) \\ \hline
    $\mathbb{E}$[$x$] & 0.612 &  0.747 &  0.747 & 0.747 \\ \hline
    $ \mathbb{E}[x^2]-\mathbb{E}[x]^2 $ &  0.2806 & 0.414 & 0.345 & 0.345 \\ \hline  
   
    \end{tabular}
    \end{center}
\end{scriptsize}
\vspace{3 mm}
\caption{Illustrative Example 4 - Mean and variance of nonlinear transformation of an external noise}
\label{tab_2}
\end{table}

\section{Uncertainty Propagation Through Nonlinear Stochastic Dynamical Systems} \label{sec_propagation}

In the previous section, we addressed the nonlinear uncertainty transformation problem where we described the moments of nonlinear transformations of uncertainties in terms of the mixed-trigonometric-polynomial moments of the original uncertainties. In this section, we address the uncertainty propagation problem through nonlinear stochastic dynamical systems described in Section \ref{sec_formulation}. More precisely, given the nonlinear stochastic dynamical system in \eqref{sys}, this section is concerned with the problem of computing statistical moments of the uncertain states $\mathbf{x}(k)$ over the planning horizon,  i.e., $m_{x^\alpha}(k)=\mathbb{E}[x^\alpha(k)]$, $\alpha \in \mathbb{N}^n$, $k=1,...,N$.

To this end, we provide two approaches including 1) \textit{Direct} and 2) \textit{Recursive} approaches. In the direct approach, we describe the moments of the uncertain states at time $k$ in terms of the moments of the initial states and the external disturbances. On the other hand, in the recursive approach, we describe the moments of the uncertain states at time $k+1$ recursively in terms of the moments of the uncertain states at the previous time step $k$. Such recursive formulation allows us to address the uncertainty propagation over the \textit{long} planning horizons.

\subsection{Direct Approach} \label{sec_direct}
In the direct method, we use the results of Section \ref{sec_UTrans} to describe the moments of the probability distributions of the uncertain states at time $k$ in terms of the moments of the initial uncertain states and the process noises. For this purpose, given the stochastic dynamical system in \eqref{sys}, we can describe the states $\mathbf{x}(k)$ in terms of the uncertainties and the given control input $\mathbf{u}^*$ by recursion of the dynamical system \eqref{sys} as follows:
\begin{equation} \label{sys_direct}
\mathbf{x}(k)=f^{[k]} \left(\mathbf{x}(0),\mathbf{u}^*(j)|_{j=0}^{k-1},\mathbf{\omega}(j)|_{j=0}^{k-1}\right)    \end{equation}

\noindent where, $f^{[k]}=f^{[k-1]} \circ f$ is the $k$-th order function composition of $f$. According to \eqref{sys_direct}, $\mathbf{x}(k)$ is a nonlinear transformation of the initial uncertainties and the process noises. Hence, we can describe the moments of $\mathbf{x}(k)$, i.e., $m_{x^{\alpha}}(k)=\mathbb{E}[x^{\alpha}(k)]$, $\alpha \in \mathbb{N}^n$, in terms of the moments of $\mathbf{x}(0)$ and $\mathbf{\omega}(j)|_{j=0}^{k-1}$ as shown in Section \ref{sec_UTrans}.\\

\textbf{Illustrative Example 5:} Consider the following stochastic nonlinear system:
\begin{align} 
\begin{split}
x(k+1)&=x(k)+v(k)cos\left(\theta(k)\right) \\
\theta(k+1)&=\theta(k)+\omega_{\theta}(k)
\notag
\end{split}
\end{align}
\noindent where, $(x,\theta)$ are the states, $v$ is the control input, and $\omega_{\theta}$ is the external disturbance. The initial states
are uncertain as 
$x(0) \sim \mathcal{U}niform(-0.1,0.1)$ and $\theta(0) \sim \mathcal{N}(0,1)$. Also, the external disturbance at each time step has Gamma distribution as $\omega_{\theta}(k) \sim \Gamma(1,2)$. Given the control input\begin{small} $v^*(k)|_{k=0}^{N-1}$\end{small}, we aim at finding the moments of the state $x$ at the end of the planning horizon, i.e., $k=N$. To this end, we first describe the state $x$ at the time step $k=N$ as follows:
\begin{align} 
\begin{split}
x(N) &=x(0)+\sum_{k=0}^{N-1}  v(k)cos\left( \theta(0)+\sum_{i=0}^{N-1}\omega_{\theta}(i) \right) \notag
\end{split}
\end{align}

Then, by taking the expected value of powers of $x$ and expanding the $cos(.)$ function, we can compute the moments of  $x(N)$, i.e., $m_{x^{^{\alpha}}}(N)=\mathbb{E}[x^{\alpha}(N)], \alpha \in \mathbb{N}$, in terms of the given control input, polynomial moments of $x(0)$, and trigonometric moments of $\theta(0)$ and $\omega_{\theta}(k)|_{k=0}^{N-1}$. 
For example, we obtain the moments of order $\alpha=1,...,5$ of $x(6)$ for the given control input $v^*(k)=0.5|_{k=0}^{5}$ as \begin{small}$[0.2974,0.7028,0.5581,1.3158,1.536]$\end{small}. 
One can verify the results using the extensive Monte Carlo simulation.\\



\textit{Remark 2}: In the direct approach, $f^{[k]}$ in \eqref{sys_direct} is a function of $k+1$ random vectors including $ \mathbf{x}(0)$ and $\mathbf{\omega}(j)|_{j=0}^{k-1}$. Hence, for large values of $k$ computation of the powers of $ f^{[k]}$, i.e., $x^{\alpha}(k)=\left( f^{[k]} \right)^{\alpha}$, $\alpha \in \mathbb{N}^n$, becomes computationally challenging. As a results, direct approach is suitable for short planning horizons and small moment orders $\alpha$. 
To address this problem, in the next section, we provide a recursive approach to compute the moments of the uncertain states over the long planning horizons.

\subsection{Recursive Approach}
In this section, we provide a recursive formulation to obtain the moments of the uncertain states of the stochastic nonlinear dynamical system in \eqref{sys}. For this purpose, we describe the moments of the probability distributions of the states at time step $k+1$ in terms of the moments of the probability distributions of the states and external disturbances at time step $k$. More precisely, we construct a new deterministic dynamical system that governs the time evolution of the moments of the states of the stochastic nonlinear system in \eqref{sys} as follows:
\begin{equation} \label{sys_mom}
    \mathbf{m}_{x^{\alpha}}(k+1)=f_{mom_{\alpha}}\left(\mathbf{m}_{x^{\alpha}}(k),\mathbf{m}_{\omega^{\alpha}}(k),\mathbf{u}(k)\right)
\end{equation}
where, $\mathbf{m}_{x^{\alpha}}$ and $\mathbf{m}_{\omega^{\alpha}}$ are the vector of all moments of order $\alpha$ of the uncertain state vector $\mathbf{x}$ and uncertain vector $\omega$, respectively.
According to \eqref{sys_mom}, moments at time step $k+1$, i.e., $\mathbf{m}_{x^{\alpha}}(k+1)$, can be described in terms of the moments at time step $k$, i.e., $\mathbf{m}_{x^{\alpha}}(k)$ and $\mathbf{m}_{\omega^{\alpha}}(k)$. Hence, given the initial moments of the uncertain states, i.e.,$\mathbf{m}_{x^{\alpha}}(0)$, and the moments of the external uncertainties, i.e., $\mathbf{m}_{\omega^{\alpha}}(k)|_{k=0}^{N-1}$, and the nominal control inputs $\mathbf{u}^*(k)|_{k=0}^{N-1}$, one can recursively compute the moments of the uncertain states over the planning horizon $k=1,...,N$.

To construct the moment dynamical system in \eqref{sys_mom}, we need the expected values of the powers of the function $f$ in dynamical system \eqref{sys}, i.e., \begin{small}
 ${m}_{x^{\alpha}}(k+1)=\mathbb{E}[x^{\alpha}(k+1)]=\mathbb{E}[\prod_{i=1}^n x_i^{\alpha_i}(k+1)]=\mathbb{E}[f^{\alpha}(\mathbf{x}(k),\mathbf{\omega}(k),\mathbf{u}(k))]$, $\alpha \in \mathbb{N}^n$\end{small}. However, note that, due to nonlinearity of function $f$,  $x^{\alpha}(k+1)$ is described in terms of monomials and also mixed trigonometric polynomial functions of the state vector $\mathbf{x}(k)$. This implies that, for a given $\alpha \in \mathbb{N}^n$, moment ${m}_{x^{\alpha}}(k+1)$ is not only a function of the polynomial moments but also the mixed-trigonometric-polynomial moments of the states. Hence, besides the update equation of the polynomial moments as in \eqref{sys_mom}, we need to obtain the update equation of such mixed-trigonometric-polynomial moments of the states as well.

To show this, we began by showing how the dynamical system for the time evolution of the first order moment of the uncertain state $x$ in illustrative example 5 can be obtained.
By substituting the dynamics of $x(k+1)$ in and applying the linearity of the expectation, we get the dynamics of the first order moment as follows:
\begin{equation} \label{sys_mom_exa_1}
\mathbb{E}[x(k+1)]=\mathbb{E}[x(k)]+ v(k)\mathbb{E}[cos(\theta(k))]
\end{equation}

By replacing the polynomial moment of $x(k)$, moment dynamical system \eqref{sys_mom_exa_1} reads as
\begin{equation} \label{sys_mom_exa_2}
m_x(k+1)=m_x(k)+ v(k)\mathbb{E}[cos(\theta(k))] \end{equation}
To fully describe the moment dynamical system in \eqref{sys_mom_exa_2} in terms of the moments, we obtain the update equation of trigonometric moment $\mathbb{E}[cos(\theta(k))]$ as follows:
\begin{align} \label{sys_mom_exa_3}
\mathbb{E} &  [cos(\theta(k+ 1))]=\mathbb{E}\left[cos\left(\theta(k)+\omega_{\theta}(k)\right)\right]
 =\mathbb{E}\left[cos(\theta(k))\right]\mathbb{E}\left[cos(\omega_{\theta}(k) ) \right] 
 - \mathbb{E}\left[sin(\theta(k)) \right] \mathbb{E}\left[sin(\omega_{\theta}(k) )\right] 
\end{align}
By replacing the trigonometric moments, \eqref{sys_mom_exa_3} reads as
\begin{align} \label{sys_mom_exa_4}
\begin{split}
m_{c_{\theta}}(k+1)= m_{c_{\omega_{\theta}}}(k) m_{c_{\theta}}(k)
 - m_{s_{\omega_{\theta}}}(k) m_{s_{\theta}}(k) \
\end{split}
\end{align}
\noindent where, 
\begin{small}  $m_{c_{{\theta}}}(k)$ \end{small} and \begin{small} $ m_{s_{{\theta}}}(k)$\end{small} are the trigonometric moments of $\theta(k)$
and \begin{small}  $m_{c_{\omega_{\theta}}}(k)$ \end{small} and \begin{small} $ m_{s_{\omega_{\theta}}}(k)$\end{small} are the trigonometric moments of the external uncertainty $\omega_{\theta}(k)$. We can compute the moments \begin{small}  $m_{c_{\omega_{\theta}}}(k)$ \end{small} and \begin{small} $ m_{s_{\omega_{\theta}}}(k)$\end{small} using the characteristic function of \begin{small} $\omega(k)$\end{small}  as shown in Section \ref{sec_mom}. 
Now, to complete the update equation in \eqref{sys_mom_exa_4}, we need to obtain the update equation of the term\begin{small} $m_{s_{{\theta}}}(k)$\end{small}. Similar to \eqref{sys_mom_exa_3} and \eqref{sys_mom_exa_4}, we can compute the update equation of \begin{small}$m_{s_{{\theta}}}(k)$\end{small} as follows:
\begin{align} \label{sys_mom_exa_5}
\begin{split}
m_{s_{\theta}}  (k+1)=
 m_{c_{\omega_{\theta}}}(k) m_{s_{\theta}}(k)
 + m_{s_{\omega_{\theta}}}(k) m_{c_{\theta}}(k)  \
\end{split}
\end{align} 

Hence, according to
\eqref{sys_mom_exa_2}, \eqref{sys_mom_exa_4}, and \eqref{sys_mom_exa_5}, we can describe the time evolution of the first order moment $m_x(k)$ in terms of the slack moment states
$m_{c_{\theta}}$ and $m_{s_{\theta}}$ as follows: 
\begin{equation} \label{sys_mom_exa_6}
\begin{bmatrix}
m_x(k+1)\\
m_{c_{\theta}}(k+1)\\
m_{s_{\theta}}(k+1)
\end{bmatrix}=
\begin{bmatrix}
1 & v(k) & 0\\
0 & m_{c_{\omega_{\theta}}}(k) & -m_{s_{\omega_{\theta}}}(k)\\
0 & m_{s_{\omega_{\theta}}}(k) & m_{c_{\omega_{\theta}}}(k)
\end{bmatrix}
\begin{bmatrix}
m_x(k)\\
m_{c_{\theta}}(k)\\
m_{s_{\theta}}(k)
\end{bmatrix}
\end{equation} 

According to the moment dynamical system in \eqref{sys_mom_exa_6}, given i) the initial moment vector \begin{small}$[m_x(0),m_{c_{\theta}}(0),m_{s_{\theta}}(0)]^T$\end{small} obtained using the probability distribution of the initial states\begin{small} $(x(0),\theta(0))$\end{small}, ii) trigonometric moments \begin{small}
 $m_{c_{\omega_\theta}}(k)|_{k=0}^{N-1}$
\end{small}  and \begin{small} $m_{s_{\omega_\theta}}(k)|_{k=0}^{N-1}$
\end{small} obtained using the probability distributions of the external uncertainties \begin{small}$\omega_{\theta}(k)|_{k=0}^{N-1}$\end{small}, and iii) control inputs \begin{small}$v(k)|_{k=0}^{N-1}$\end{small}, we can recursively update the moment $m_x(k)$ over the planning horizon \begin{small}$k=0,...,N$\end{small}.

Similarly, we can obtain the dynamics of the higher order moments of $x(k)$. For this purpose, similar to \eqref{sys_mom_exa_6}, we need to identify the slack moment states. This process, however, is tedious and is easily subject to human error, especially for larger moment order $\alpha$.  
We have addressed this issue in the risk assessment problem for Dubins vehicle by developing the $\textit{TreeRing}$ algorithm that uses a dependency graph to identify all the slack moment states \cite{mom_risk_2,TreeRing_2}.  

In this paper, we use a different approach and provide a general framework to construct the moment-state dynamical system in \eqref{sys_mom}. The main idea is to transform the nonlinear stochastic dynamical system in \eqref{sys} into a new augmented linear-state system. In this case, due to the linear relation of the states of the augmented linear-state system at time $k$ and $k+1$, polynomial moments of order $\alpha$ of the states at time $k+1$ can be described \textit{only} in terms of the polynomial moments of order $\alpha$ of the states at time $k$. Hence, we do not need to look for a set of slack moment states as described in (21)-(26)

To show how the proposed algorithm works, in the following, we define the augmented linear-state system for the stochastic nonlinear dynamical system in \eqref{sys}. We then show how one can use the obtained augmented linear-state system to construct the moment-state dynamical systems that describe the time evolution of the moments of the states of the original stochastic nonlinear dynamical system in \eqref{sys}. \\

\textbf{Augmented Linear-State System:} 
We define the augmented linear-state system for
nonlinear stochastic system \eqref{sys} as follows
\begin{equation} \label{sys_aug}
\mathbf{x}_{aug}(k+1)=A_k(\mathbf{\omega},\mathbf{u})\mathbf{x}_{aug}(k)    
\end{equation}
where $\mathbf{x}_{aug}$ is the augmented state vector that is defined in terms of i) states of the original nonlinear stochastic system in \eqref{sys} and ii) nonlinear functions of these states appeared in function $f$ of dynamics \eqref{sys}. Elements of matrix $A_k$ is defined in terms of, in general, mixed trigonometric polynomial functions of external disturbance $\omega(k)$ and control input $u(k)$.\\ 

\textit{Remark 3}: Note that although the augmented system in \eqref{sys_aug} is linear in terms of the augmented states $\mathbf{x}_{aug}$, it is still nonlinear in terms of the external disturbance $\mathbf{\omega}(k)$ and the control input $\mathbf{u}(k)$.\\



For example, we obtain the augmented linear-state system for the stochastic nonlinear dynamical system of illustrative example 5 as follows. We initially define the augmented state vector in terms of state $x(k)$ and the nonlinear state term of the dynamical system as \begin{small}$\mathbf{x}_{aug}(k)=[x(k),cos(\theta(k))]^T$\end{small}\footnote{Since we are interested in the moments of state $x$, we do not consider state $\theta$ directly in $\mathbf{x}_{aug}$.}. By doing so, 
the augmented system is obtained as follows:
\begin{equation} 
\mathbf{x}_{aug}(k+1)=
\begin{bmatrix}
1 & v(k) \\
0 & {cos(\omega_{\theta}}(k))
\end{bmatrix}
\mathbf{x}_{aug}(k)-\begin{bmatrix}
0\\
sin(\omega_{\theta}(k))sin(\theta(k))
\end{bmatrix} \notag
\end{equation} 
Since the obtained augmented system is still nonlinear, in the next step, we update the augmented state vector $\mathbf{x}_{aug}$ by adding the nonlinear state term of the obtained augmented system as\begin{small} $\mathbf{x}_{aug}(k)=[x(k),cos(\theta(k)),sin(\theta(k))]^T$\end{small}. By doing so, the augmented linear-state system for the nonlinear stochastic system of illustrative example 5 is obtained as follows: 
\begin{equation}\label{exa5_aug}
\mathbf{x}_{aug}(k+1)=
\begin{bmatrix}
1 & v(k) & 0\\
0 &  cos(\omega_{\theta}(k)) & -sin(\omega_{\theta}(k))\\
0 & sin(\omega_{\theta}(k)) & cos(\omega_{\theta}(k))
\end{bmatrix}\mathbf{x}_{aug}(k) 
\end{equation} 
The obtained augmented linear-state system is equivalent to the original nonlinear stochastic system. \\

\textit{Remark 4}: The process of constructing augmented linear-state system \eqref{sys_aug} for nonlinear stochastic system \eqref{sys}
is similar to the \textit{Kerner} and \textit{Carleman} linearization methods. Using these linearization techniques, one can 
transform deterministic nonlinear ordinary differential systems into linear systems by introducing suitable new variables. For more information see \cite{Kerner_1, Carleman_1}.\\

We use augmented linear-state system \eqref{sys_aug} to obtain moment-state linear systems that govern the time evolution of the moments of the uncertain states of nonlinear stochastic system \eqref{sys} as follows:\\


\textbf{Moment-State Linear System:
} We define the moment-state linear system for augmented linear-state system \eqref{sys_aug} as follows:
\begin{equation} \label{sys_mom_rec}
    \mathbf{m}_{x^{\alpha}}(k+1)=A_{mom_{\alpha}}(k)\mathbf{m}_{x^{\alpha}}(k)
\end{equation}
where, $\mathbf{m}_{x^{\alpha}}$ is the vector of all moments of order $\alpha$ of the augmented state vector $\mathbf{x}_{aug}$ and $A_{mom_{\alpha}}$ is a matrix in terms of the control input $\mathbf{u}(k)$ and mixed-trigonometric-polynomial moments of the external disturbance $\omega(k)$. To construct the moment-state linear system in \eqref{sys_mom_rec}, we need the expected values of the monomials of order $\alpha$ of the vector $\mathbf{x}_{aug}(k+1)$, e.g.,
$\mathbf{m}_{x^{\alpha}}(k+1)=\mathbb{E}[x_{aug}^{\alpha}(k+1)]$. Since $\mathbf{x}_{aug}(k+1)$ is a linear function of $\mathbf{x}_{aug}(k)$ as in \eqref{sys_aug}, we can describe the polynomial moments of order $\alpha$ of $\mathbf{x}_{aug}(k+1)$ completely in terms of the polynomial moments of order $\alpha$ of $\mathbf{x}_{aug}(k)$. Hence, we do not need to look for a set of slack moment states as described in \eqref{sys_mom_exa_1}-\eqref{sys_mom_exa_6}.


For example, we obtain the moment-state linear system of order $\alpha=1$ of the form $ \mathbf{m}_{x}(k+1)=A_{mom_{1}}(k)\mathbf{m}_{x}(k)$
for the augmented linear-state system of illustrative example 5 in \eqref{exa5_aug} as follows:
\begin{equation}
\mathbf{m}_{x}(k+1)=
\begin{bmatrix}
1 & v(k) & 0\\
0 & m_{c_{\omega_{\theta}}}(k) & -m_{s_{\omega_{\theta}}}(k)\\
0 & m_{s_{\omega_{\theta}}}(k) & m_{c_{\omega_{\theta}}}(k)
\end{bmatrix}\mathbf{m}_{x}(k)
\notag 
\end{equation} 
\noindent where, $\mathbf{m}_{x}(k)=\left[ \ \mathbb{E}[x(k)],\  \mathbb{E}[cos(\theta(k))],\  \mathbb{E}[sin(\theta(k))]  \  \right ]^T$ is the vector of all moments of order $\alpha=1$ of $\mathbf{x}_{aug}(k)$. 
Also, matrix $A_{mom_{1}}$ is described in terms of the control input $v(k)$ and the first order trigonometric moments of the external uncertainty $\omega_{\theta}(k)$. Note that the moment vector $\mathbf{m}_{x}$ includes the first order moment of the uncertain state of the original stochastic nonlinear system in illustrative example 5, i.e., $\mathbb{E}[x]$, and also two slack moment states.

Similarly, we obtain the moment-state linear system of order $\alpha=2$ of the form $\mathbf{m}_{x^{2}}(k+1)=A_{mom_{2}}(k)\mathbf{m}_{x^{2}}(k)
$ where $\mathbf{m}_{x^2}(k)$ is the vector of all moments of order
$\alpha=2$ of $\mathbf{x}_{aug}(k)$ as follows:

\begin{small}\begin{equation}
\mathbf{m}_{x^2}(k) = \left[ \
\mathbb{E}[x^2(k)], \ 
\mathbb{E}[x(k)cos(\theta(k))], \ 
\mathbb{E}[x(k)sin(\theta(k))], 
\mathbb{E}[cos^2(\theta(k))], \
 \mathbb{E}[cos(\theta(k))sin(\theta(k))], \
\mathbb{E}[sin^2(\theta(k))] \   \right]^T \notag
\end{equation} 
\end{small}
Also, matrix $A_{mom_{2}}$ is obtained in terms of the control input $v(k)$ and the first and second order trigonometric moments of the external uncertainty $\omega_{\theta}(k)$ as follows:
\begin{center}
    \resizebox{0.7\linewidth}{!}{%
$A_{mom_{2}}(k)=\begin{bmatrix}
 1&     2v(k)&         0&              v^2(k)&                       0&                  0\\
 0& m_{c_{\omega_\theta}}(k)& -m_{s_{\omega_\theta}}(k)&       v(k)m_{c_{\omega_\theta}}(k)&            -v(k)m_{s_{\omega_\theta}}(k)&                  0\\
 0& m_{s_{\omega_\theta}}(k)&  m_{c_{\omega_\theta}}(k)&       v(k)m_{s_{\omega_\theta}}(k)&             v(k)m_{c_{\omega_\theta}}(k)&                  0\\
 0&        0&         0&        m_{c^2_{\omega_\theta}}(k)&    -2m_{c_{\omega_\theta}s_{\omega_\theta}}(k)&         m_{s^2_{\omega_\theta}}(k)\\
 0&        0&         0& m_{c_{\omega_\theta}s_{\omega_\theta}}(k)& m_{c^2_{\omega_\theta}}(k) - m_{s^2_{\omega_\theta}}(k)& -m_{c_{\omega_\theta}s_{\omega_\theta}}(k)\\
 0&        0&         0&        m_{s^2_{\omega_\theta}}(k)&     2m_{c_{\omega_\theta}s_{\omega_\theta}}(k)&         m_{c^2_{\omega_\theta}}(k)\\
 \end{bmatrix} 
$ }
\end{center}

\textit{Remark 5}: We can construct the deterministic linear moment-state systems of the form \eqref{sys_mom_rec} for different moment order $\alpha$ in the offline step. Hence, we can use the obtained moment systems to propagate the moments of the uncertain initial states over the planning horizon in real-time. 
More precisely, given initial moments $\mathbf{m}_{x^{\alpha}}(0)$ and $A_{mom_{\alpha}}(k)|_{k=0}^{N-1}$, the moments at time step $N$ can be obtained by recursion of  
$ \mathbf{m}_{x^{\alpha}}(k+1)=A_{mom_{\alpha}}(k)\mathbf{m}_{x^{\alpha}}(k), k=0,...,N-1$. Similarly, we can describe the moments by the solution of the linear moment system as $ \mathbf{m}_{x^{\alpha}}(N)=\Pi_{k=0}^{N-1}A_{mom_{\alpha}}(k) \mathbf{m}_{x^{\alpha}}(0)$.\\

The following result holds true.\\

\textbf{Theorem 1:} There exist finite moment-state linear systems of the form \eqref{sys_mom_rec} that describe the exact time evolution of the moments of order $\alpha \in \mathbb{N}^{n}$ of the uncertain states of nonlinear stochastic dynamical system \eqref{sys} if and only if there exists an equivalent finite augmented linear-state system of the form \eqref{sys_aug}.\\

\textit{Proof}: The process of constructing augmented system \eqref{sys_aug} by introducing a set of new states, transforms a \textit{general} class of nonlinear stochastic dynamical systems into equivalent linear systems with infinite number of states, e.g., similar to \textit{Kerner} and \textit{Carleman} linearization methods. In this case, infinite dimensional moment-state linear systems of the form \eqref{sys_mom_rec} describe the exact time evolution of the moments of the obtained infinite dimensional augmented linear system. Also, marginal moment states of such moment dynamical systems describe the exact time evolution of the moments of the uncertain states of the original stochastic nonlinear system. To obtain finite (approximate) moment-linear systems, one needs to truncate the augmented state vector and work with an approximate truncated augmented linear system. 

In order to have finite exact moment-state linear systems, we need a finite exact augmented linear system of the form \eqref{sys_aug}. Such augmented system exists, if and only if, there exists a finite augmented state vector $\mathbf{x}_{aug}$ containing the states of the original nonlinear stochastic system and a finite set of slack states such that the mapping between $\mathbf{x}_{aug}(k+1)$ and $\mathbf{x}_{aug}(k)$ is linear as in \eqref{sys_aug}.  \hfill  $\square$ \\

\textit{Remark 6}: In the absence of exact finite moment-state linear system \eqref{sys_mom_rec}, one can use the direct method for exact moment propagation described in Section \ref{sec_direct}.\\

In Section \ref{sec_results}, we will obtain the exact finite moment-state linear systems to describe the exact time evolution of the moments of the uncertain states of nonlinear stochastic autonomous and robotic systems.

\section{Experiments} \label{sec_results}

In this section, we benchmark our uncertainty propagation method on different autonomous and robotic systems and show how the exact moments of the uncertain states can be computed over the planning horizon. More precisely, in this section, we address the uncertainty propagation problem for ground, underwater, and  aerial  vehicles,  robotic  arms  and  walking robots. To this end, we obtain the \textit{exact} finite moment-state linear systems of the form \eqref{sys_mom_rec}. We use such moment-state systems, in a receding horizon fashion, to propagate the initial uncertainties and obtain the exact moments of the uncertain states over the planing horizon in \textit{real-time}. 
One can use the obtained exact moments of the uncertain states to improve the planning and control under uncertainty and nonlinear estimation problems.
 
\subsection{Underwater Vehicle}
\begin{figure}[t]
    \centering
    \includegraphics[scale=0.47]{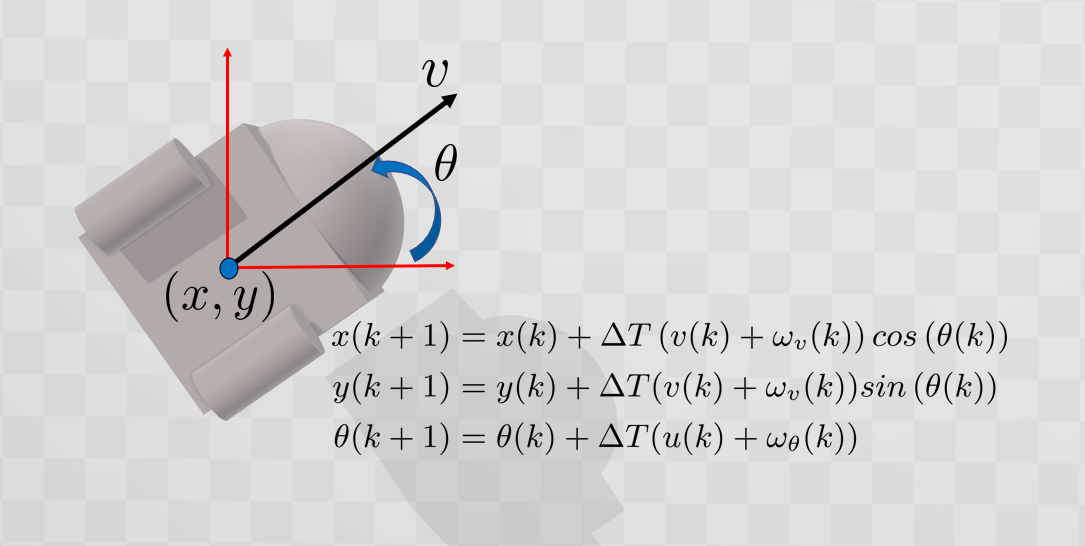}
    \caption{Example A- Uncertain Underwater Vehicle Model}
    \label{fig_A_UW_1}
\end{figure}

The motion of an underwater vehicle in the presence of external disturbances is modeled as in Figure \ref{fig_A_UW_1}, \cite{Exa_Under}. In this model, $(x,y,\theta)$ are the position and orientation of the vehicle, $(v,u)$ are the linear and angular velocities, $(\omega_v,\omega_{\theta})$ are the external disturbances, and $\Delta T=0.1$ is the sampling time step. At each time step, external disturbances have uniform distributions over $[-0.1,0.1]$. 
Initial states are also uncertain as \begin{small}$x(0)\sim \mathcal{U}niform(-0.1,0.1)$, $y(0)\sim \mathcal{U}niform(-0.1,0.1)$,\end{small}
and \begin{small}$\theta(0)\sim \mathcal{U}niform(\pi/4-0.1,\pi/4+0.1)$\end{small}. 

\begin{figure}[b]
    \centering
    \includegraphics[scale=0.20]{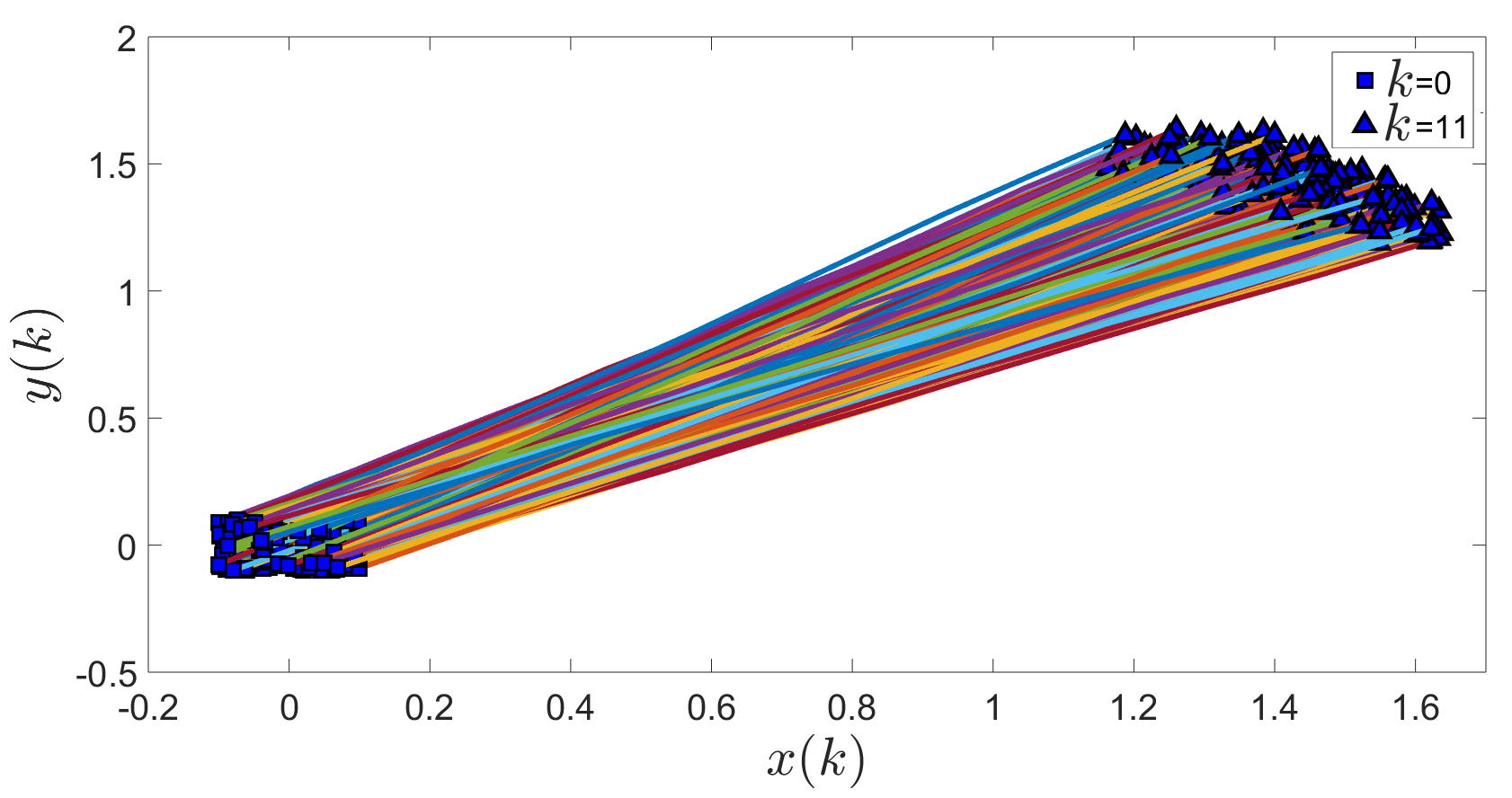}
    \caption{Example A- Trajectories of the uncertain underwater vehicle for the different realization of probabilistic  initial states and external disturbances over the planning horizon. }
    \label{fig_A_UW_2}
\end{figure}

Given the control inputs \begin{small}$(v^*(k)=2,u^*(k)=0)|_{k=0}^{10}$\end{small}, we aim at finding the moments of the future position of the vehicle along the planning horizon. Figure \ref{fig_A_UW_2} shows the uncertain behaviour of the underwater vehicle in the presence of uncertain initial states and external uncertainties. To obtain the moments of the position  along the planning horizon, we first obtain the equivalent augmented linear-state system of the form $\mathbf{x}_{aug}(k+1)=A_k(\omega_{\theta}, \omega_{v},u,v)\mathbf{x}_{aug}(k)$ where $\mathbf{x}_{aug}=[ x, y, cos(\theta), sin(\theta)]^T$ is the augmented state vector and matrix $A_k(\omega_{\theta}, \omega_{v},u,v)$ reads as

\begin{center}
    \begin{footnotesize}
    \resizebox{0.6\linewidth}{!}{%
$A_k(\omega_{\theta}, \omega_{v},u,v)=\begin{bmatrix}  1 & 0 &  \Delta T (v(k)+\omega_{v}(k)) & 0\\ 0 & 1 & 0 & \Delta T (v(k)+\omega_{v}(k)) \\ 0 & 0 & cos\left(\Delta T (u(k) +  \omega_{\theta}(k))\right) & -sin \left(\Delta T( u(k) + \omega_{\theta}(k))\right)\\ 0 & 0 & sin\left(\Delta T (u(k) + \omega_{\theta}(k))\right) &
cos\left(\Delta T (u(k) +  \omega_{\theta}(k))\right)
 \\\end{bmatrix} $}
\end{footnotesize}
\end{center}

\begin{figure}[h]
    \centering
    \includegraphics[scale=0.31]{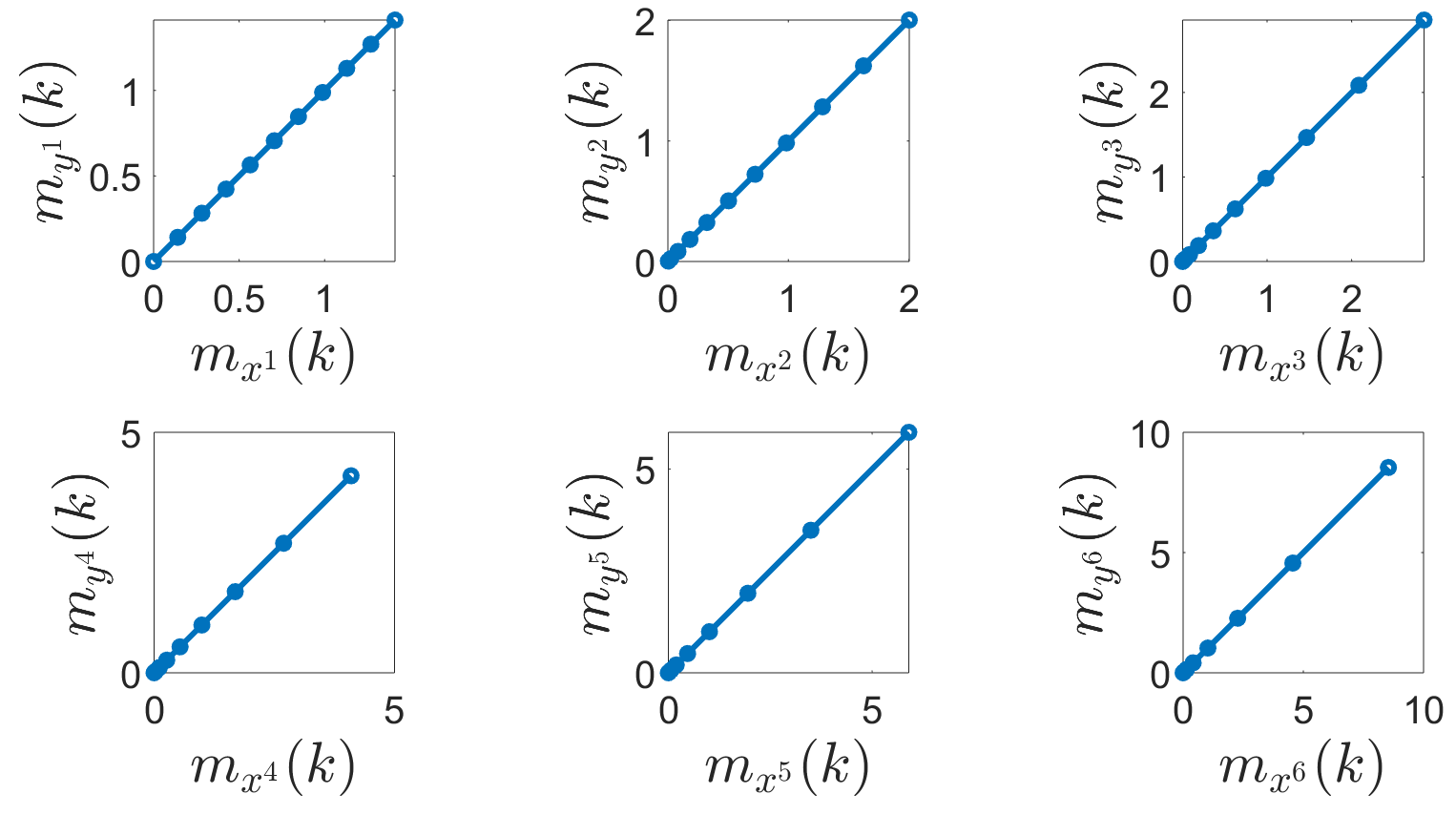}
    \caption{Example A- Moments of the uncertain position of the underwater vehicle over the planning horizon, including $\left( \mathbb{E}[x^{\alpha}(k)],\mathbb{E}[y^{\alpha}(k)] \right),\  \alpha=1,...,6, \ k=0,...,11.$}
    \label{fig_A_UW_3}
\end{figure}

Using the equivalent augmented linear-state system, we obtain the exact moment-state linear systems of the form \eqref{sys_mom_rec} for moment orders $\alpha=1,...,6$. Using the obtained moment-state linear systems, we compute the moments of order $\alpha=1,...,6$ of the position along the planning horizon. Figure \ref{fig_A_UW_3} shows the obtained moments \begin{small}$\left( \mathbb{E}[x^{\alpha}(k)],\mathbb{E}[y^{\alpha}(k)] \right),\  \alpha=1,...,6, \ k=0,...,11.$ \end{small} 
One can verify the results using the extensive Monte Carlo simulation.

\subsection{Ground Vehicle}
\begin{figure}[h]
    \centering
    \includegraphics[scale=0.39]{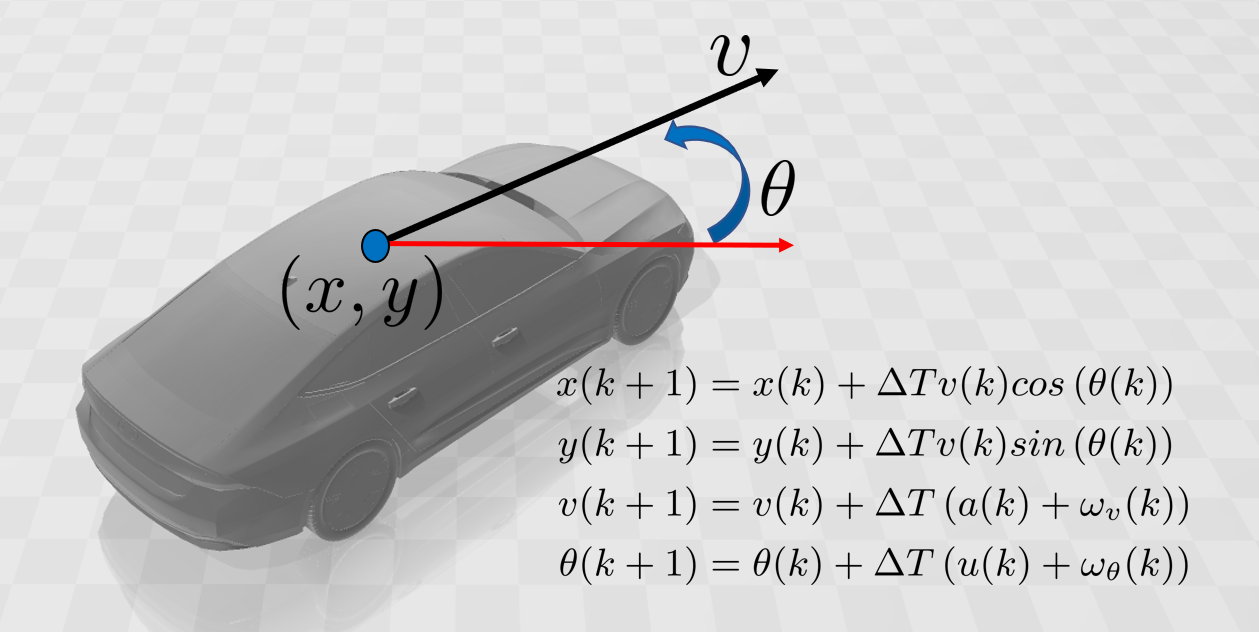}
    \caption{ Example B- Uncertain Ground Vehicle Model}
    \label{fig_B_GV_1}
\end{figure}

The motion of a ground vehicle in the presence of external disturbances is modeled as in Figure \ref{fig_B_GV_1},  \cite{robot_Linear}. In this model, $(x,y,v,\theta)$ are the position, linear velocity, and orientation of the vehicle, $(a,u)$ are the linear acceleration and angular velocity, $(\omega_v,\omega_{\theta})$ are the external disturbances, and $\Delta T=0.1$ is the sampling time step. Probability distributions of the external disturbances are \begin{small}$\omega_{v}(k) \sim
\mathcal{N}(0,1)$\end{small} and 
\begin{small}$\omega_{\theta}(k) \sim \mathcal{B}eta(1,3)$\end{small}. Initial states are also uncertain as  \begin{small}$x(0)\sim \mathcal{U}{niform}(-0.1,0.1)$, $y(0)\sim\mathcal{U}{niform}(-0.5,0.5)$, $v(0)\sim\mathcal{U}{niform}(0,0.1)$\end{small}, and \begin{small}$\theta(0)\sim\mathcal{U}{niform}(\pi/2-0.1,\pi/2+0.1)$\end{small}.

\begin{figure}[h]
    \centering
    \includegraphics[scale=0.2]{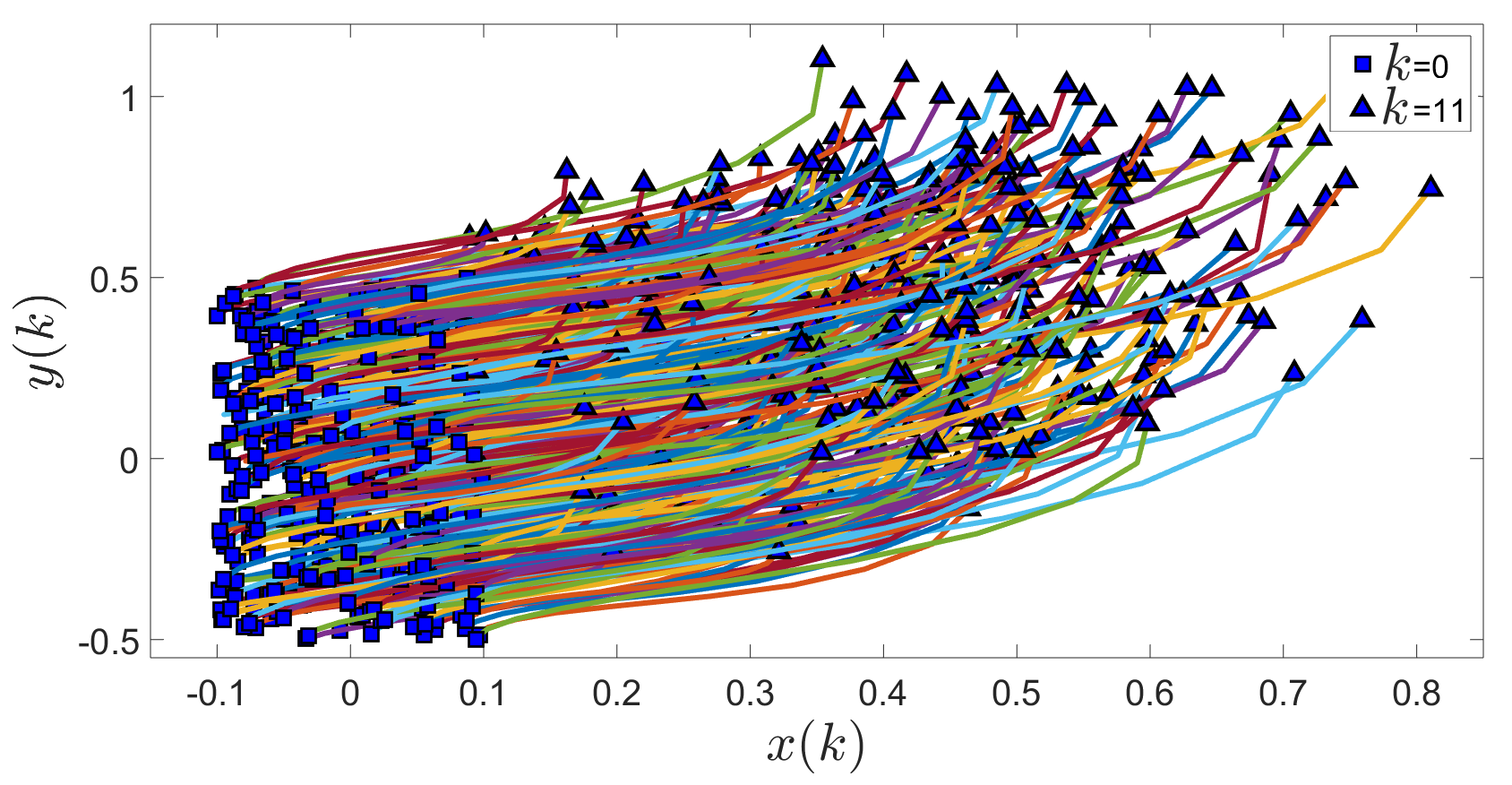}
    \caption{Example B- Trajectories of the uncertain ground vehicle for the different realization  of  probabilistic  initial  states  and  external  disturbances over  the planning horizon.}
    \label{fig_B_GV_2}
\end{figure}

Given the control inputs \begin{small}$\left(a^*(k)=1,u^*(k)=\frac{2\pi}{7.5}(k-5)\right)|_{k=0}^{10}$\end{small}, we aim at finding the moments of the future position of the vehicle along the planning horizon. Figure  \ref{fig_B_GV_2}  shows  the uncertain behaviour of the ground vehicle in the presence of uncertain initial states and external uncertainties.
To obtain the moments of the position  along the planning horizon, we first obtain the equivalent augmented linear-state system of the form $\mathbf{x}_{aug}(k+1)=A_k(\omega_{\theta}, \omega_{v},a,u)\mathbf{x}_{aug}(k)$ where $\mathbf{x}_{aug}=[ x,y,vcos(\theta), vsin(\theta), cos(\theta), sin(\theta)]^T$ is the augmented state vector and matrix $A_k(\omega_{\theta}, \omega_{v},a,u)$ reads as

\begin{center}
    \resizebox{0.95\linewidth}{!}{%
$A_k(\omega_{\theta}, \omega_{v},a,u)=\begin{bmatrix}  1 & 0 &  \Delta T & 0 & 0 & 0 \\ 1 & 0 &  0 & \Delta T & 0 & 0  \\
0 & 0 & cos(\Delta T (u + \omega_{\theta})) & -sin(\Delta T ( u +  \omega_{\theta})) & \Delta T (a+\omega_{v})cos(\Delta T (u + \omega_{\theta}))  & -\Delta T (a+\omega_{v})sin(\Delta T (u + \omega_{\theta}))\\ 
0 & 0 & sin(\Delta T (u + \omega_{\theta})) & cos(\Delta T ( u +  \omega_{\theta})) & \Delta T (a+\omega_{v})sin(\Delta T (u + \omega_{\theta}))  & \Delta T (a+\omega_{v})cos(\Delta T (u + \omega_{\theta}))\\
 0 & 0 & 0 & 0& cos(\Delta T (u +  \omega_{\theta})) & -sin(\Delta T (u + \ \omega_{\theta}))\\ 0 & 0 & 0 & 0 & sin(\Delta T (u + \omega_{\theta})) &
cos(\Delta T ( u +  \omega_{\theta}))
 \\\end{bmatrix} $}

\end{center}

\begin{figure}[h]
    \centering
    \includegraphics[scale=0.22]{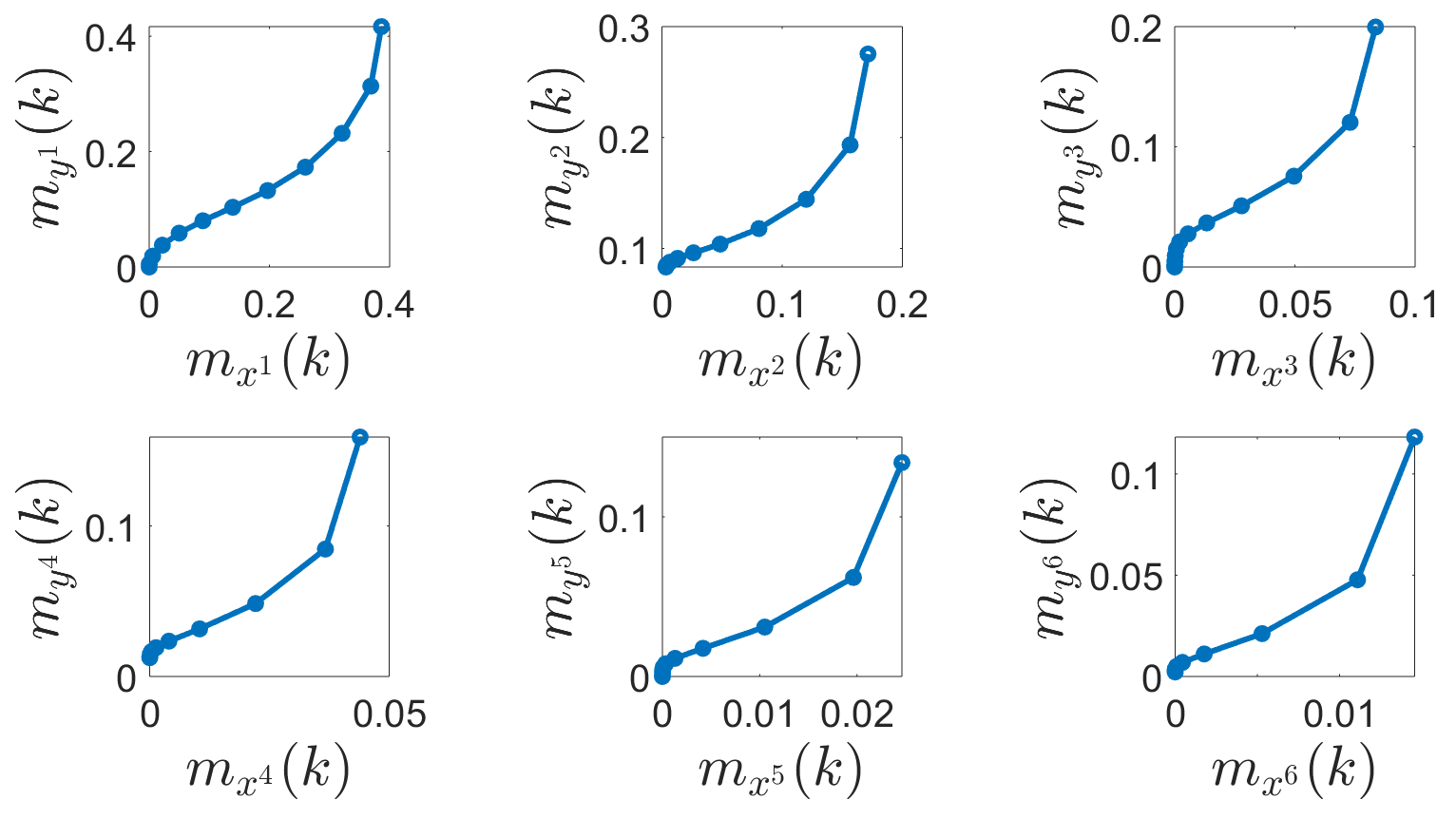}
\caption{Example B- Moments of the uncertain position of the ground vehicle over the  planning  horizon, including $\left( \mathbb{E}[x^{\alpha}(k)],\mathbb{E}[y^{\alpha}(k)] \right),\  \alpha=1,...,6, \ k=0,...,11.$}
    \label{fig_B_GV_3}
    \end{figure}

Using the equivalent augmented linear-state system, we obtain the exact moment-state linear systems of the form \eqref{sys_mom_rec} for moment orders $\alpha=1,...,6$. Using the obtained moment-state linear systems, we compute the moments of order $\alpha=1,...,6$ of the position along the planning horizon. Figure \ref{fig_B_GV_3} shows the obtained moments \begin{small}$\left( \mathbb{E}[x^{\alpha}(k)],\mathbb{E}[y^{\alpha}(k)] \right),\  \alpha=1,...,6, \ k=0,...,11.$\end{small} One can verify the results using the extensive Monte Carlo simulation.

\subsection{Walking Robot}
\begin{figure}[h]
    \centering
    \includegraphics[scale=0.33]{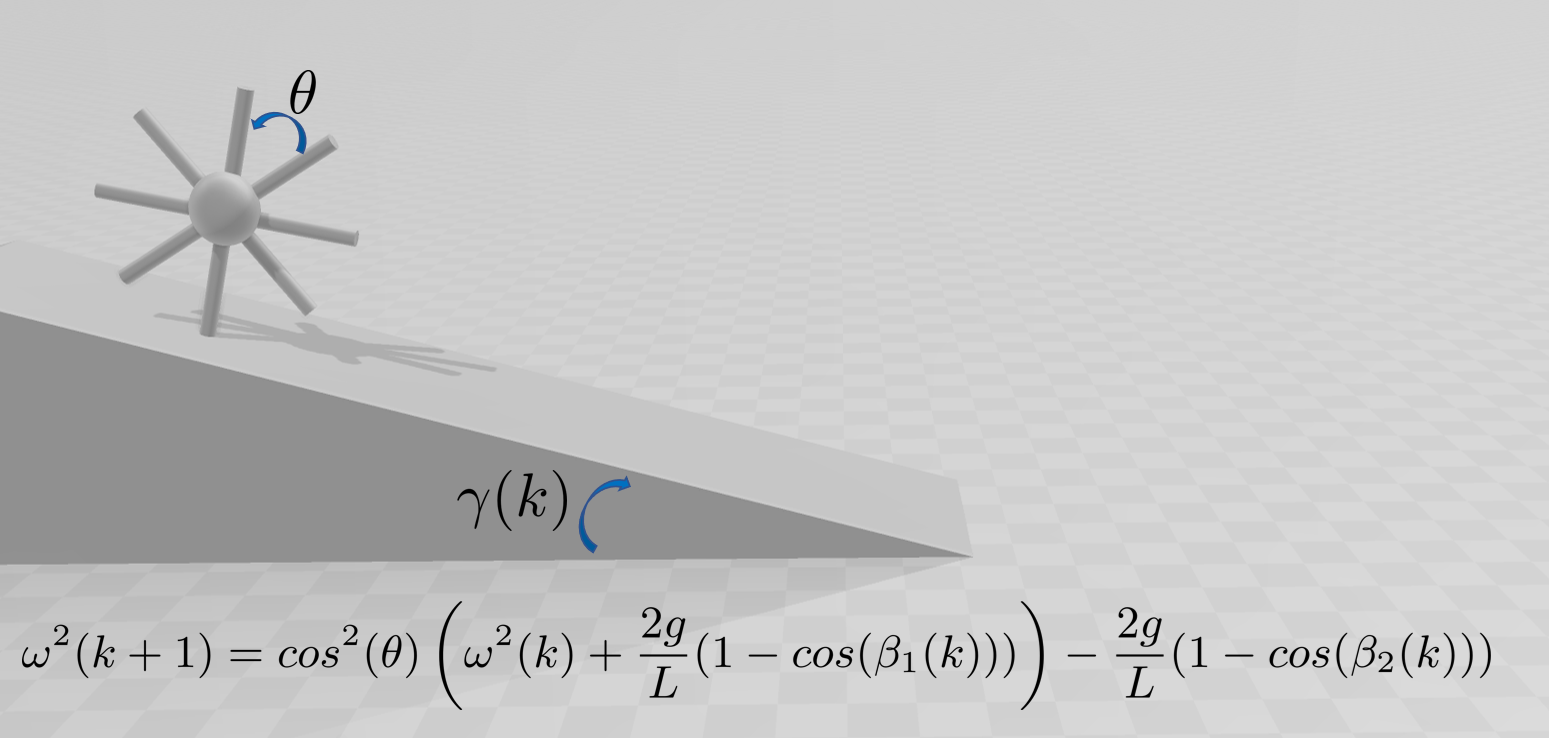}
    \caption{Example C- Uncertain Rimless Wheel Model}
    \label{fig_C_RV_1}
\end{figure}

The motion of a walking robot on rough terrain is modeled by a rimless wheel on a ramp with a random slope angle. Rimless wheel has $n_s=8$ spokes of length $L=0.5$ where the angle between the spokes is $\theta=\frac{2\pi}{n_s}$. Motion of the rimless wheel rolling down the ramp is modeled as in Figure \ref{fig_C_RV_1}, \cite{Exa_Walk_1, Exa_Walk_2, Exa_Walk_3}.

In this model, $\omega(k)$ is the angular velocity of the angle between the stance leg and the slope normal, 
and $\beta_1=\frac{\theta}{2}+\gamma$, $\beta_2=\frac{\theta}{2}-\gamma$. The uncertain slope angle is modeled by \begin{small}$\gamma(k) \sim \mathcal{N}(\pi/4,0.5)$\end{small}. The initial state is also uncertain as \begin{small}$\omega^2(0) \sim \mathcal{U}niform(-0.1,0.1)$\end{small}. Given the probability distributions of the initial state \begin{small}$\omega^2(0)$\end{small} and $\gamma(k)|_{k=0}^{10}$, we aim at finding the moments of $\omega^2(k)$ over the planning horizon. We use the original linear motion dynamics to obtain the update equation of the moments. The motion dynamics is linear in terms of the state $\omega^2$ as \begin{small}$\omega^2(k+1)=A\omega^2(k)+b(k)$\end{small} where 
\begin{small} $A=cos^2(\frac{2\pi}{n_s})$ \end{small} and
\begin{small}
$b(k)=cos^2(\frac{2\pi}{n_s})\frac{2g}{L}\left(1-cos\left(\frac{\pi}{n_s}+\gamma(k)\right)\right) - \frac{2g}{L}\left(1-cos\left(\frac{\pi}{n_s}-\gamma(k)\right)\right)$\end{small} is the nonlinear transformation of the uncertain parameter $\gamma$. 
The obtained update equation of the moment of order $\alpha \in \mathbb{N}$ is as follows:
\begin{equation}
m_{\omega^{2\alpha}}(k+1)=\mathbb{E}[  \omega ^ {2\alpha}(k+1)]=\mathbb{E} \left[ \left( A\omega^2(k)+b(k)\right)^{\alpha} \right] =\sum_{i=0}^{\alpha} \binom{\alpha}{i} \mathbb{E}\left[b^{\alpha-i}(k)\right]A^i\mathbb{E}\left[\omega^{2i}(k)\right]  =\sum_{i=0}^{\alpha} \binom{\alpha}{i} m_{b^{\alpha-i}}(k) A^im_{\omega^{2i}}(k) \nonumber
\end{equation}
where, $m_{\omega^{2i}}$ is the moment of order $i$ of state $\omega^2$. Also, $m_{b^{\alpha-i}}$ is the moment of order $\alpha-i$ of the uncertain parameter $b$ and can be computed  in terms of the trigonometric moments of $\gamma$ as described in \ref{sec_UTrans}. Note that in the obtained moment update equation, moment of order $\alpha$ at time $k+1$,  i.e., $m_{\omega^{2\alpha}}(k+1)$, is a linear function of the sequence of the  moments of orders $i=0,...,\alpha$ at time $k$, i.e., $m_{\omega^{2i}}(k), i=0,...,\alpha$.
We use the obtained moment update equation to construct the linear moment-state system to propagate the moments of orders $\alpha=0,1,2$ as follows: 
\begin{equation}
\begin{bmatrix}
m_{\omega^{0}}(k+1)\\
m_{\omega^{2}}(k+1)\\
m_{\omega^{4}}(k+1)
\end{bmatrix}=
\begin{bmatrix}
1 & 0 & 0\\
m_{b}(k) & A & 0\\
m_{b^2}(k) & 2m_{b}(k)A & A^2
\end{bmatrix}
\begin{bmatrix}
m_{\omega^{0}}(k)\\
m_{\omega^{2}}(k)\\
m_{\omega^{4}}(k)
\end{bmatrix} \notag
\end{equation}
Propagated moments over the planing horizon are computed as \begin{small}$m_{\omega^0}(k)=1|_{k=0}^{11}$
\end{small}, \begin{small}$m_{\omega^2}=[  0 ,   2.76,    4.14,    4.83,    5.18,     5.35,   5.44,  5.48, \\ 5.50 ,   5.51,    5.52 ]$\end{small}, and \begin{small}
$ m_{\omega^4}=[0.003,   74.5,  100.7,  111.1,  115.6,  117.7,  118.7,  119.2,  119.4,  119.6,  119.6].$\end{small} One can verify the results using the extensive Monte Carlo simulation.

\subsection{Planar  Aerial Vehicle}
\begin{figure}[h]
    \centering
    \includegraphics[scale=0.33]{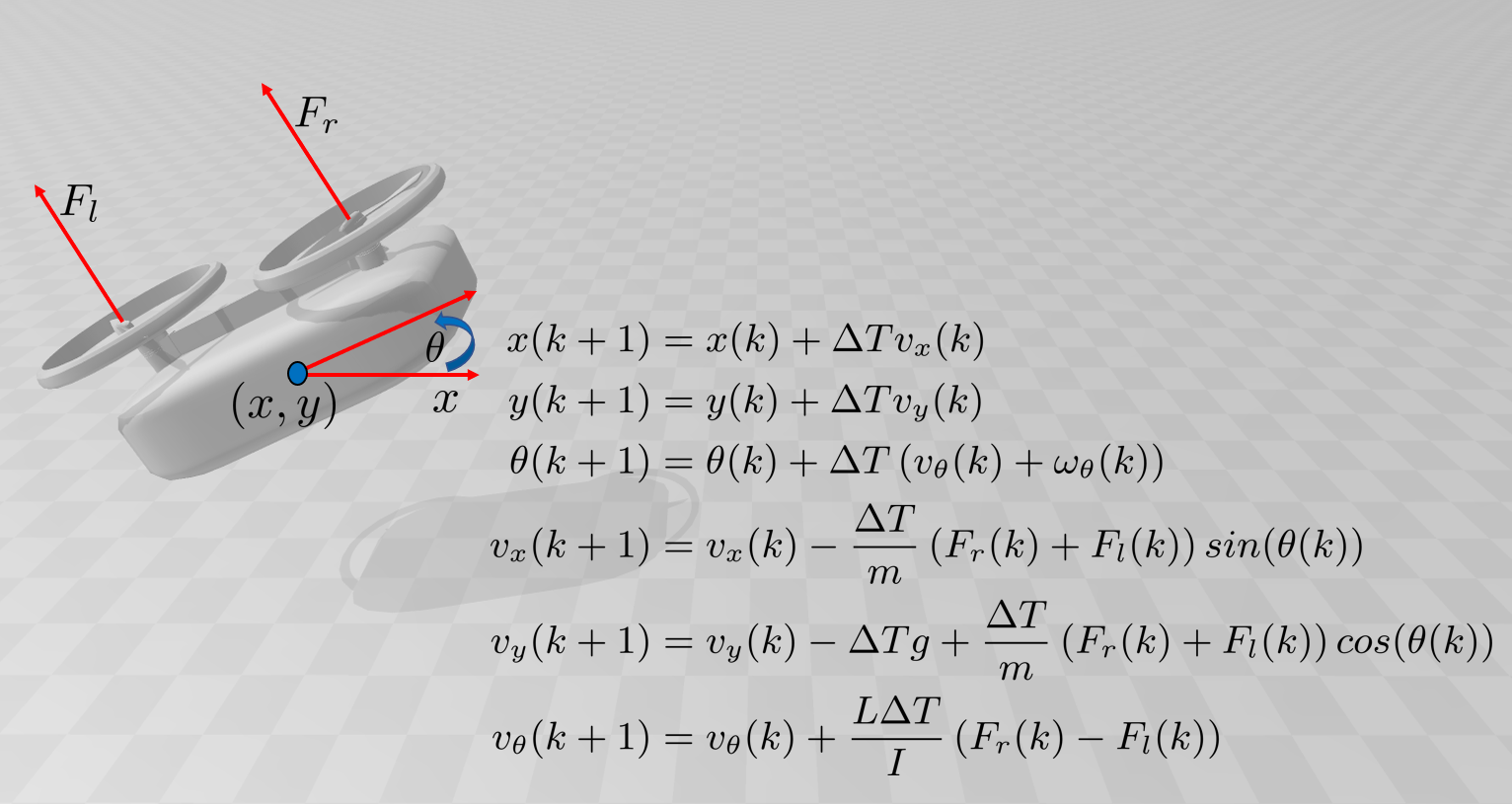}
    \caption{Example D- Uncertain Planar Model of an Aerial Vehicle}
    \label{fig_D_A2V_1}
\end{figure}

The vertical and horizontal motions of an aerial vehicle in the presence of wind disturbances is modeled as in Figure \ref{fig_D_A2V_1}, \cite{Exa_Walk_3}.
In this model, $(x,y)$ are the horizontal and vertical positions, $\theta$ is the attitude, $(v_x,v_y)$ are the linear velocities, $v_{\theta}$ is the angular velocity, $(F_l,F_r)$ are the forces generated by the left and right rotors. Also, $m=0.5$ is the mass,   $L=0.25$ is the length of the rotor arm,  $I = 0.00383$ is the moment of inertia, $g = 9.8$ is the acceleration due to gravity, and $\Delta T=0.1$ is the sampling time step. Wind disturbance is modeled by \begin{small}$\omega_{\theta}(k) \sim \mathcal{U}niform(-0.1,0.1)$\end{small}. Initial states are also uncertain as \begin{small}$x(0)\sim \mathcal{U}niform(-0.1,0.1)$, $y(0)\sim \mathcal{U}niform(0.4,0.5)$, $\theta(0)\sim \mathcal{U}niform(-0.1,0.1)$, $v_x(0)\sim \mathcal{U}niform(-0.1,0.1)$, $v_y(0)\sim \mathcal{U}niform(-0.1,0.1)$\end{small}. 

\begin{figure}[h]
    \centering
    \includegraphics[scale=0.202]{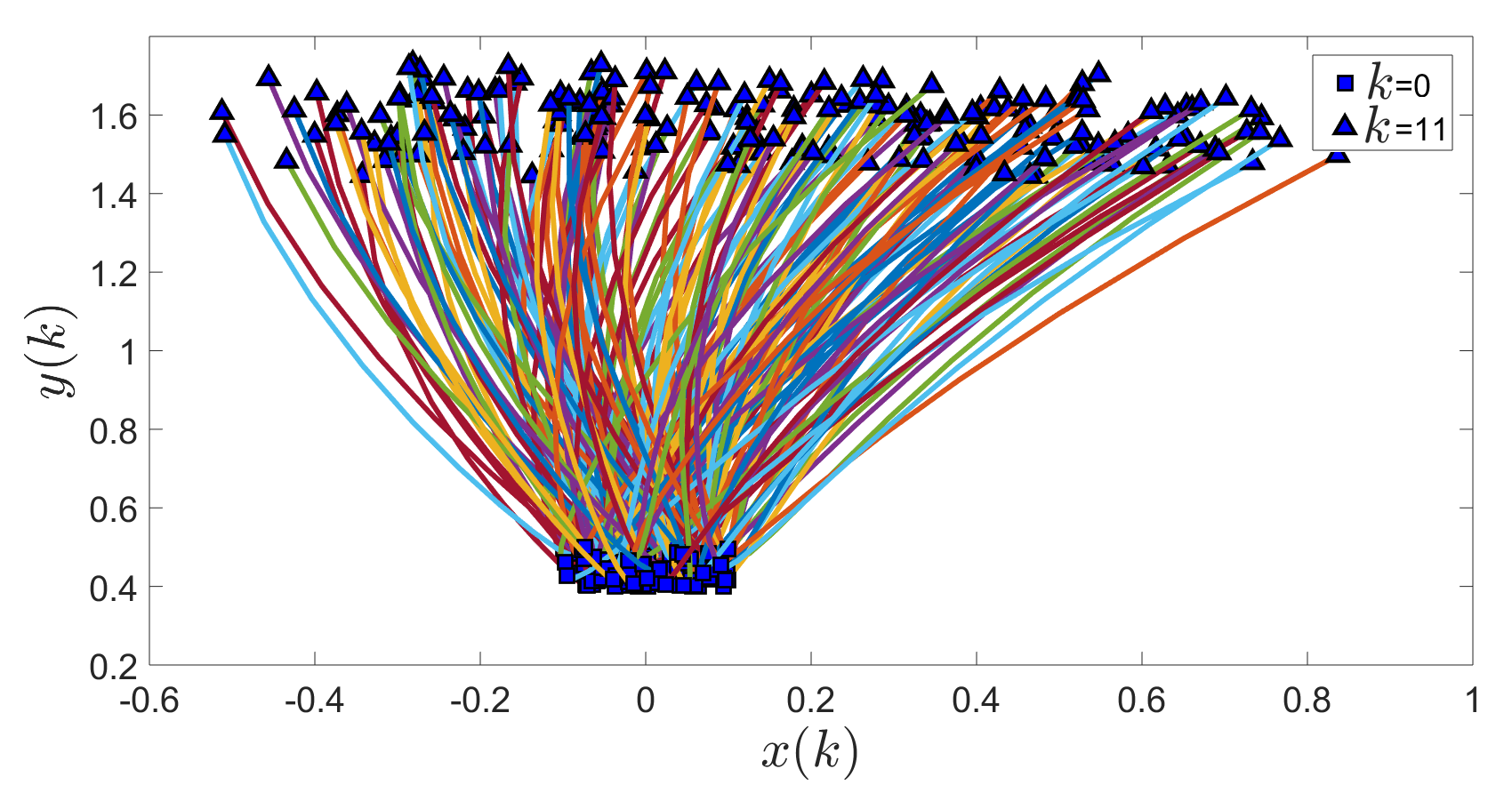}
    \caption{Example  D-  The vertical-horizontal trajectories  of  the uncertain  aerial  vehicle  for  the different realization  of  probabilistic  initial  states  and  external  disturbances over  the planning horizon.}
    \label{fig_D_A2V_2}
\end{figure}

Given the control inputs \begin{small}$(F_l^*(k)=3,F_r^*(k)=3.01)|_{k=0}^{10}$\end{small}, we aim at finding the moments of the future location of the aerial vehicle over the planning horizon. Figure  \ref{fig_D_A2V_2}  shows  the uncertain behaviour of the aerial vehicle in the presence of uncertain initial states and external uncertainties.
To obtain the moments of the position  along the planning horizon, we first obtain the equivalent augmented linear-state system of the form $\mathbf{x}_{aug}(k+1)=A_k(\omega_{\theta},F_l,F_r)\mathbf{x}_{aug}(k)$ where $\mathbf{x}_{aug}=[ x,y,v_x,v_y, a_g, cos(\theta), sin(\theta)]^T$ is the augmented state vector. We define the state $a_g$ as $a_g(k+1)=a_g(k), a_g(0)=\Delta T g$. By introducing the constant state $a_g$, we can obtain the moment-state system of the form \eqref{sys_mom_rec}. In the absence of state $a_g$, moments of order $\alpha$ at time $k+1$ depends on the all moments of orders $i=0,...,\alpha$ (similar to Example C). The obtained matrix $A_k(\omega_{\theta}, F_l,F_r)$ reads as

\begin{center}
\begin{footnotesize}
    \resizebox{0.65\linewidth}{!}{%
$A_k(\omega_{\theta}, F_l,F_r)=\begin{bmatrix} 1 & 0 & \Delta T & 0 & 0 & 0 & 0  \\ 0 & 1 & 0 & \Delta T & 0 & 0 & 0  \\ 0& 0& 1 & 0&0&0&- \frac{\Delta T}{m} (F_{l}+F_{r} ) \\  0& 0& 0 & 1 & -1 & \frac{\Delta T}{m} (F_{l}+F_{r} ) & 0 \\
0 & 0 & 0 & 0 & 1 & 0 & 0\\ 
0 & 0 & 0 & 0 & 0 & cos(\Delta T (v_{\theta}+\omega_{\theta})) & -sin(\Delta T (v_{\theta}+\omega_{\theta}))
\\ 0 & 0 & 0 & 0 & 0 & sin(\Delta T (v_{\theta}+\omega_{\theta})) & cos(\Delta T (v_{\theta}+\omega_{\theta}))
  \end{bmatrix}
 $}
\end{footnotesize}
\end{center}

\begin{figure}[h]
    \centering
    \includegraphics[scale=0.21]{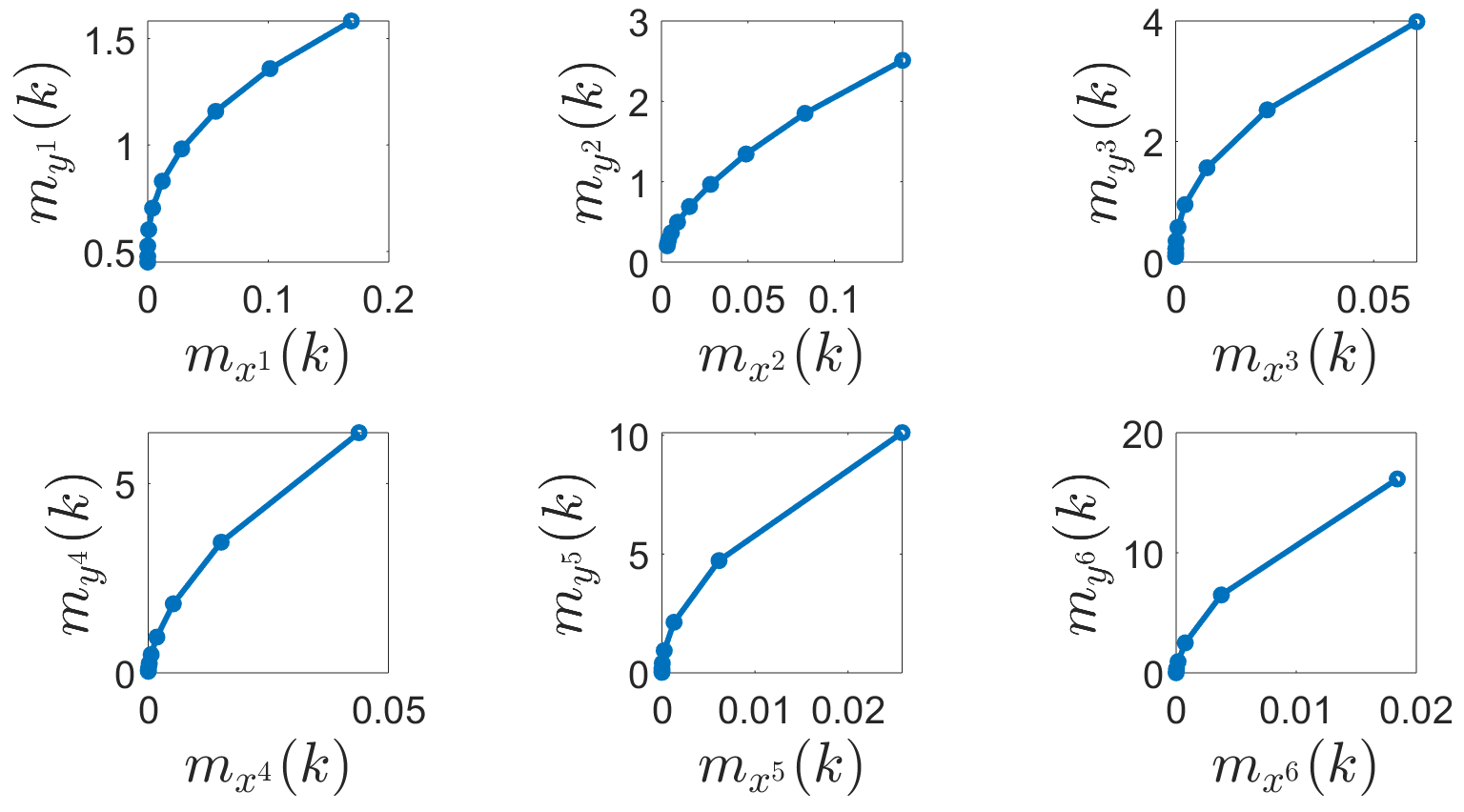}
    \caption{Example C- Moments  of  the uncertain  vertical-horizontal position  of  the aerial  vehicle  over the  planning  horizon,  including $\left( \mathbb{E}[x^{\alpha}(k)],\mathbb{E}[y^{\alpha}(k)] \right),\  \alpha=1,...,6, \ k=0,...,11.$}
    \label{fig_D_A2V_3}
\end{figure}

Using the equivalent augmented linear-state system, we obtain the exact moment-state linear systems of the form \eqref{sys_mom_rec} for moment orders $\alpha=1,...,6$. Using the obtained moment-state linear systems, we compute the moments of order $\alpha=1,...,6$ of the position along the planning horizon. Figure \ref{fig_D_A2V_3} shows the obtained moments \begin{small}$\left( \mathbb{E}[x^{\alpha}(k)],\mathbb{E}[y^{\alpha}(k)] \right),\  \alpha=1,...,6, \ k=0,...,11.$\end{small} One can verify the results using the extensive Monte Carlo simulation.

\subsection{3D Aerial Vehicle}
\begin{figure}[h]
    \centering
    \includegraphics[scale=0.32]{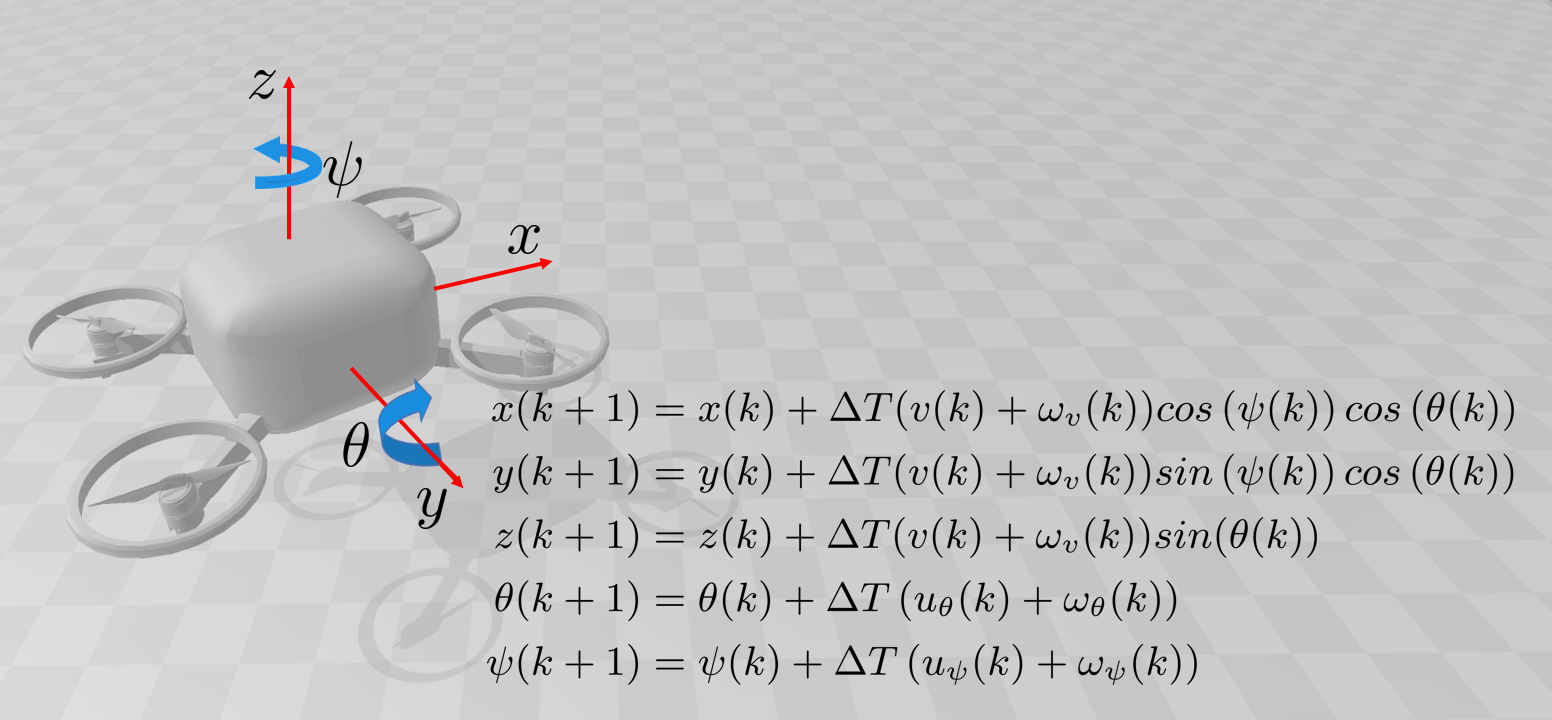}
    \caption{Example E- Uncertain Aerial Vehicle Model}
    \label{fig_G_A3V_1}
\end{figure}

The motion of an aerial vehicle in 3D space in the presence of wind disturbances is modeled as in Figure \ref{fig_G_A3V_1}, \cite{Exa_Under}. In this model, $(x,y,z)$ are the position, $(\theta,\psi)$ are the orientations around $y$ and $z$ axis, $v$ is the linear velocity, $(u_{\theta},u_{\psi})$ are the angular velocities, and $\Delta T=0.1$ is the sampling time step. Wind disturbance is modeled by \begin{small}$\omega_v(k) \sim \mathcal{B}eta(1,3)$, $\omega_{\theta}(k) \sim  \mathcal{N}(0,0.3)$, $ \omega_{\psi}(k) \sim \mathcal{U}niform(-0.1,0.1)$\end{small}. Initial states are also uncertain as \begin{small}$x(0) \sim \mathcal{U}niform(-0.1,0.1)$, $y(0) \sim \mathcal{U}niform(0.1,0.1)$, $z(0) \sim \mathcal{U}niform(0.1,0.3)$, $\theta(0) \sim \mathcal{B}eta(1,3)$, $\psi(0) \sim \mathcal{B}eta(3,3)$\end{small}.

\begin{figure}[h]
    \centering
    \includegraphics[scale=0.2]{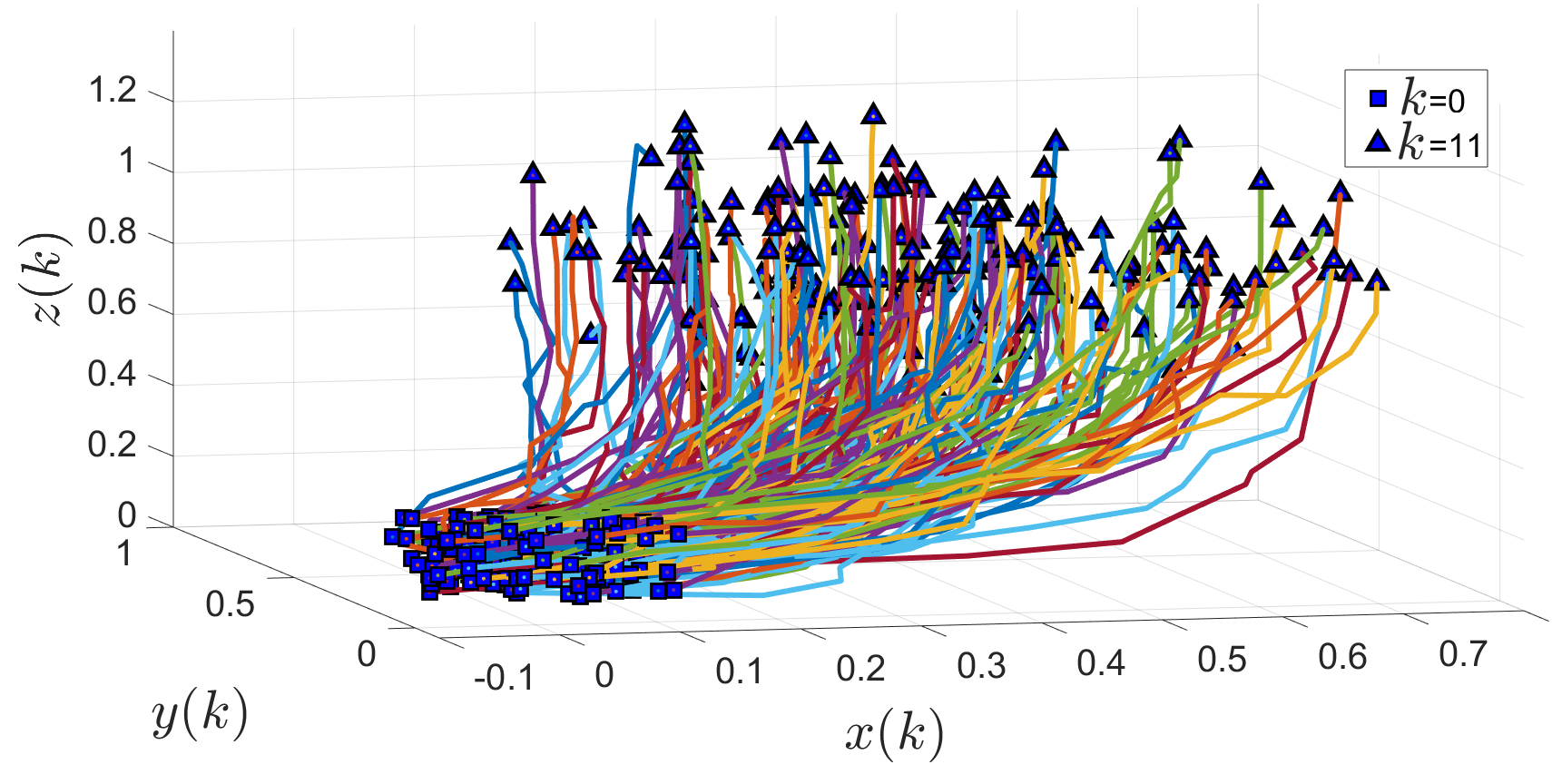}
    \caption{Example  G-  Trajectories  of  the uncertain  aerial  vehicle  for  the different realization  of  probabilistic  initial  states  and  external  disturbances over  the planning horizon.}
    \label{fig_G_A3V_2}
\end{figure}

\begin{figure}[h]
    \centering
    \includegraphics[scale=0.3]{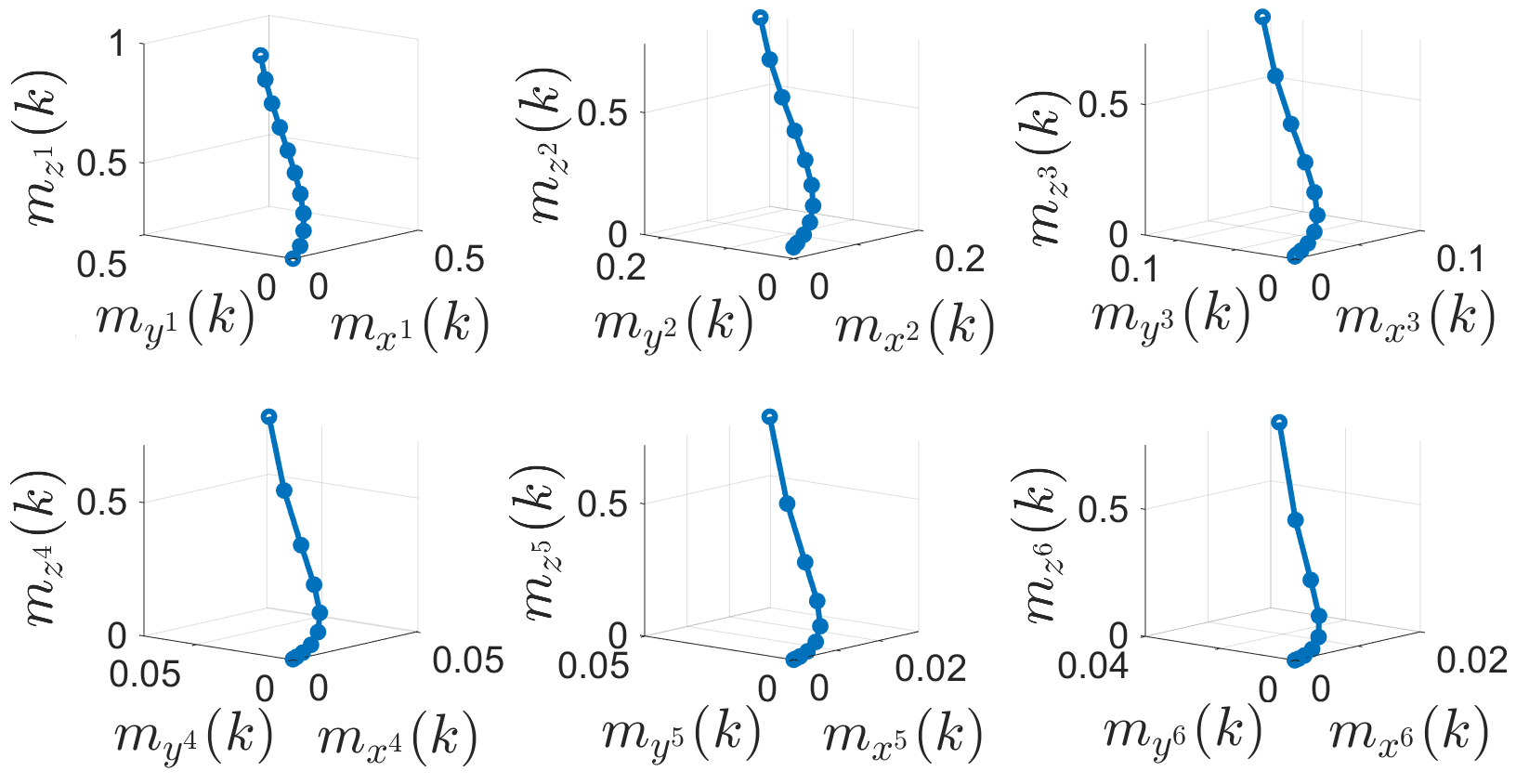}
    \caption{Example G- Moments  of  the uncertain position  of  the aerial  vehicle  over the  planning  horizon,  including $\left( \mathbb{E}[x^{\alpha}(k)],\mathbb{E}[y^{\alpha}(k)],\mathbb{E}[z^{\alpha}(k)] \right),\  \alpha=1,...,6, \ k=0,...,11.$}
    \label{fig_G_A3V_3}
\end{figure}

Given the control inputs \begin{small}$(v^*(k)=1, u_{\theta}^*(k)=1,u_{\psi}^*(k)=1)|_{k=0}^{10}$\end{small}, we aim at finding the moments of the future location of the vehicle over the planning horizon. Figure  \ref{fig_G_A3V_2}  shows  the uncertain behaviour of the aerial vehicle in the presence of uncertain initial states and external uncertainties.
To obtain the moments of the position  along the planning horizon, we first obtain the equivalent augmented linear-state system of the form $\mathbf{x}_{aug}(k+1)=A_k(\omega_{v},\omega_{\theta},\omega_{\psi}, v,u_{\theta},u_{\psi})\mathbf{x}_{aug}(k)$ where
\begin{small}$\mathbf{x}_{aug}=[ x,y,z, cos(\psi)cos(\theta),
sin(\psi)cos(\theta),
cos(\psi)sin(\theta) , sin(\psi)sin(\theta), cos(\theta),sin(\theta)]^T$\end{small}
 is the augmented state vector and matrix $A_k(\omega_{v},\omega_{\theta},\omega_{\psi}, v,u_{\theta},u_{\psi})$ reads as
\begin{center}
    \resizebox{0.99\linewidth}{!}{%
$A_k(\omega_{v},\omega_{\theta},\omega_{\psi}, v,u_{\theta},u_{\psi})=\begin{bmatrix}  1 & 0 & 0 & \Delta T (v + \omega_{v}) & 0 & 0 & 0 & 0 & 0 \\ 0 & 1 & 0 & 0 & \Delta T (v + \omega_{v})  & 0 & 0 & 0 & 0 \\ 0 & 0 & 1 & 0 & 0 & 0 & 0 & 0 & \Delta T (v + \omega_{v}) \\
0 & 0 & 0 & cos(U_1)cos(U_2) & -sin(U_1)cos(U_2) & -cos(U_1)sin(U_2) & sin(U_1)sin(U_2) & 0 & 0\\ 0 & 0 & 0 & sin(U_1)cos(U_2) &  cos(U_1)cos(U_2) & - sin(U_1)sin(U_2) & - cos(U_1)sin(U_2) & 0 & 0\\ 0 & 0 & 0 & cos(U_1)sin(U_2) & -sin(U_1)sin(U_2) & cos(U_1)cos(U_2) &-sin(U_1)cos(U_2) & 0 & 0\\ 0 & 0 & 0 & sin(U_1)sin(U_2) & cos(U_1)sin(U_2) & sin(U_1)cos(U_2) & cos(U_1)cos(U_2) & 0 & 0 \\ 0 & 0 & 0 & 0 & 0 & 0 &0 & cos(U_2) & -sin(U_2) \\ 0 & 0 & 0 & 0 & 0 & 0 &0 & sin(U_2) & cos(U_2)   \end{bmatrix}
$}
\end{center}
where $U_1=\Delta T (u_{\psi}+\omega_{{\psi}})$ and $\ U_2=\Delta T (u_{\theta}+\omega_{{\theta}})$. 
Using the equivalent augmented linear-state system, we obtain the exact moment-state linear systems of the form \eqref{sys_mom_rec} for moment orders $\alpha=1,...,6$. Using the obtained moment-state linear systems, we compute the moments of order $\alpha=1,...,6$ of the position along the planning horizon. Figure \ref{fig_G_A3V_3} shows the obtained moments \begin{small}$\left( \mathbb{E}[x^{\alpha}(k)],\mathbb{E}[y^{\alpha}(k),\mathbb{E}[z^{\alpha}(k)] \right),\  \alpha=1,...,6, \ k=0,...,11.$ \end{small} One can verify the results using the extensive Monte Carlo simulation.

\subsection{Differential-Drive Mobile Robot}
\begin{figure}[h]
    \centering
    \includegraphics[scale=0.47]{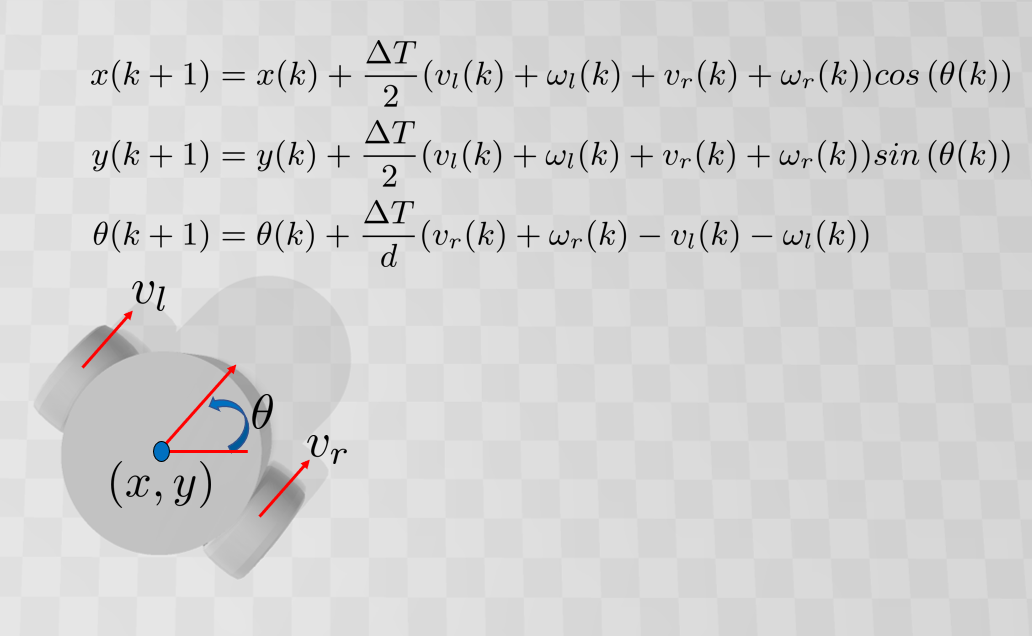}
    \caption{Example F- Uncertain Model of a Differential-Drive Mobile Robot}
    \label{fig_F_DM_1}
\end{figure}

The motion of a differential-drive mobile robot in the presence of external disturbances is modeled as in Figure \ref{fig_F_DM_1}, \cite{robot_Linear}. 
In this model, $(x,y,\theta)$ are the position and orientation, $(v_l,v_r)$ are the speed of the left and right wheels, $d=1$ is the distance between the wheels, and $\Delta T=0.1$ is the sampling time step. External disturbances are modeled with \begin{small}$\omega_l(k) \sim \mathcal{U}niform(-0.1,0.1)$, $\omega_r(k) \sim \mathcal{B}eta(1,3)$\end{small}. Initial states are also uncertain as \begin{small}$x(0) \sim \mathcal{U}niform(-0.1,0,1)$, $y(0) \sim \mathcal{U}niform(-0.1,0,1)$, $\theta(0) \sim \mathcal{N}(0,0.1)$\end{small}. 

\begin{figure}[h]
    \centering
    \includegraphics[scale=0.2]{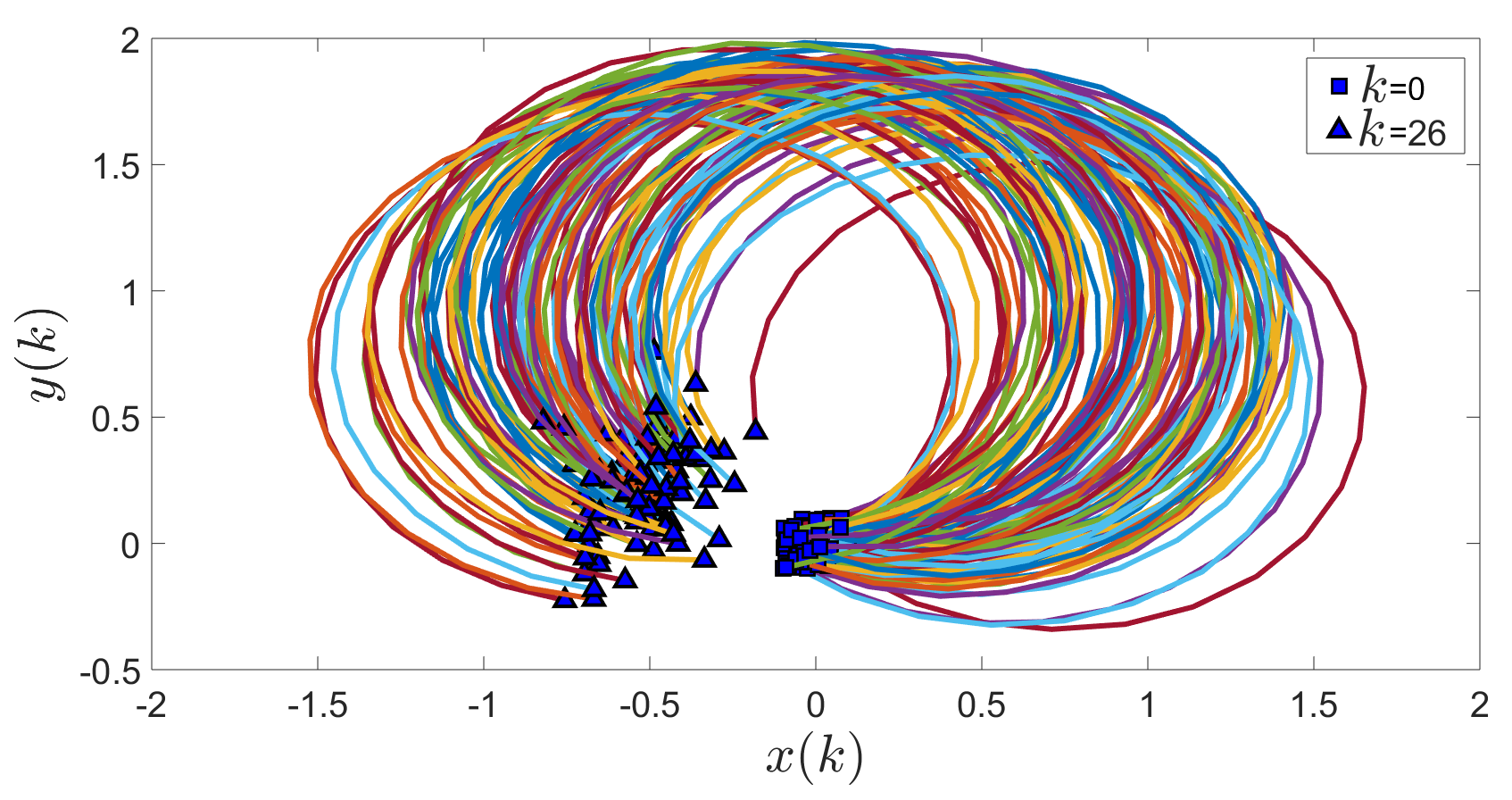}
    \caption{Example  F-  Trajectories  of  the uncertain differential-drive mobile robot for the different realization of  probabilistic initial states and  external disturbances over the planning horizon.}
    \label{fig_F_DM_2}
\end{figure}

Given the control inputs \begin{small}$(v_l^*(k)=1,v_r^*(k)=3)|_{k=0}^{25}$\end{small}, we aim at finding the moments of the future location of the mobile robot over the planning horizon. Figure \ref{fig_F_DM_2} shows the uncertain behaviour of the robot in the presence of uncertain initial states and external uncertainties. To obtain the moments of the position  along the planning horizon, we first obtain the equivalent augmented linear-state system of the form $\mathbf{x}_{aug}(k+1)=A_k(\omega_{l},\omega_{r},v_l,v_r)\mathbf{x}_{aug}(k)$ where
\begin{small}$\mathbf{x}_{aug}=[ x,y, cos(\theta),sin(\theta)]^T$\end{small}
 is the augmented state vector and matrix $A_k(\omega_{l},\omega_{r},v_l,v_r)$ reads as
\begin{center}
    \resizebox{0.7\linewidth}{!}{%
$A_k(\omega_{l},\omega_{r},v_l,v_r)=\begin{bmatrix}  1 & 0 &  \frac{\Delta T}{2} (v_{l}+\omega_{{l}} + v_{r}+\omega_{{r}}) & 0\\ 0 & 1 & 0 & \frac{\Delta T}{2} (v_{l}+\omega_{{l}} + v_{r}+\omega_{{r}}) \\ 0 & 0 & cos\left( \frac{\Delta T}{d} (v_{r}+\omega_{{r}} - v_{l}-\omega_{{l}})\right) & -sin\left( \frac{\Delta T}{d} (v_{r}+\omega_{{r}} - v_{l}-\omega_{{l}})\right)\\ 0 & 0 & sin\left( \frac{\Delta T}{d} (v_{r}+\omega_{{r}} - v_{l}-\omega_{{l}})\right) &
cos\left( \frac{\Delta T}{d} (v_{r}+\omega_{{r}} - v_{l}-\omega_{{l}})\right)
 \\\end{bmatrix}
$}
\end{center}

\begin{figure}[h]
    \centering
    \includegraphics[scale=0.21]{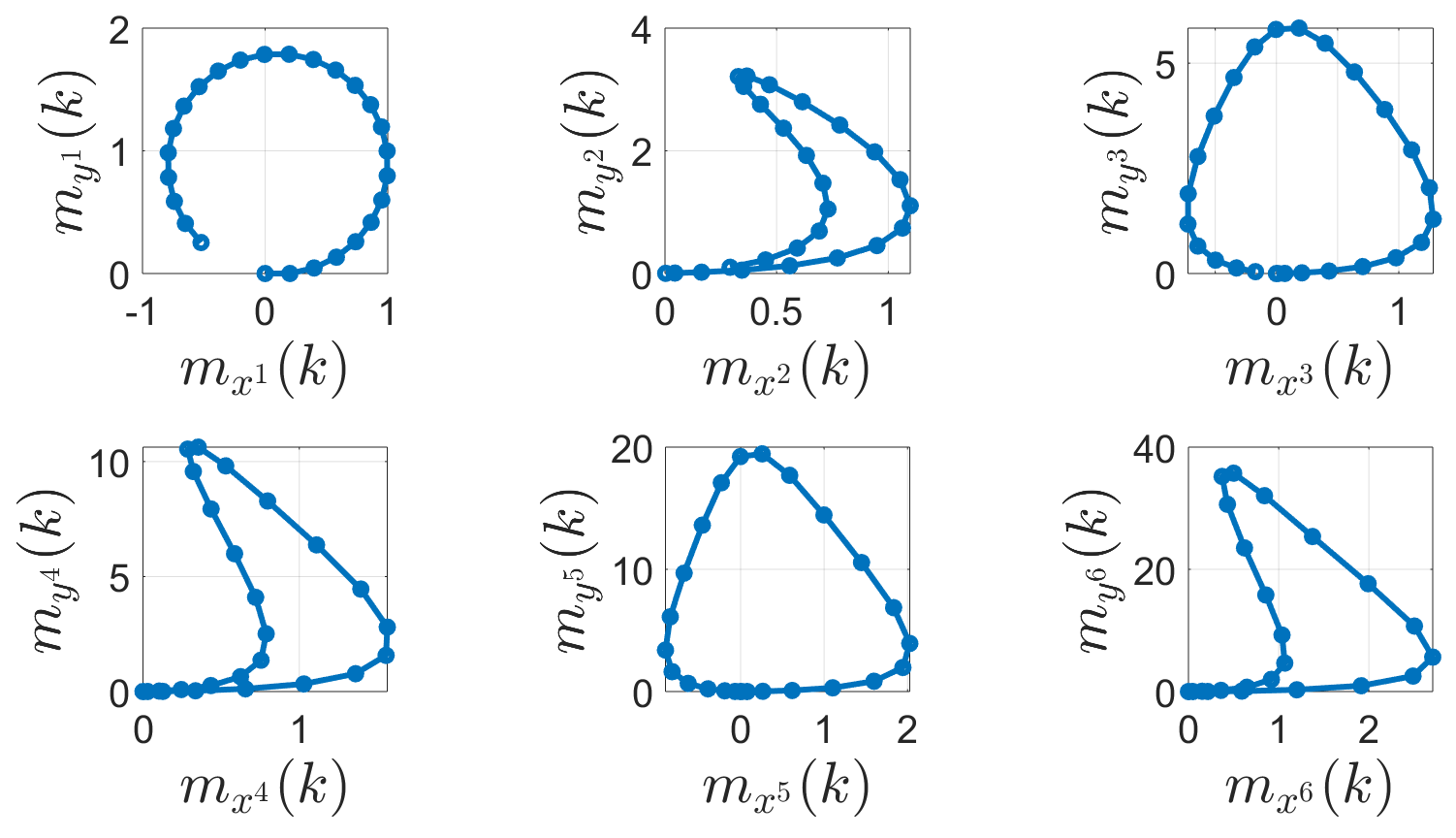}
    \caption{Example F- Moments  of  the uncertain position of the differential-drive mobile robot  over the  planning  horizon,  including $\left( \mathbb{E}[x^{\alpha}(k)],\mathbb{E}[y^{\alpha}(k)] \right),\  \alpha=1,...,6, \ k=0,...,26.$}
    \label{fig_F_DM_3}
\end{figure}

Using the equivalent augmented linear-state system, we obtain the exact moment-state linear systems of the form \eqref{sys_mom_rec} for moment orders $\alpha=1,...,6$. Using the obtained moment-state linear systems, we compute the moments of order $\alpha=1,...,6$ of the position along the planning horizon. Figure \ref{fig_F_DM_3} shows the obtained moments \begin{small}$\left( \mathbb{E}[x^{\alpha}(k)],\mathbb{E}[y^{\alpha}(k)] \right),\  \alpha=1,...,6, \ k=0,...,26.$ \end{small} One can verify the results using the extensive Monte Carlo simulation.

\subsection{Mobile Robotic Arm}
\begin{figure}[h]
    \centering
    \includegraphics[scale=0.29]{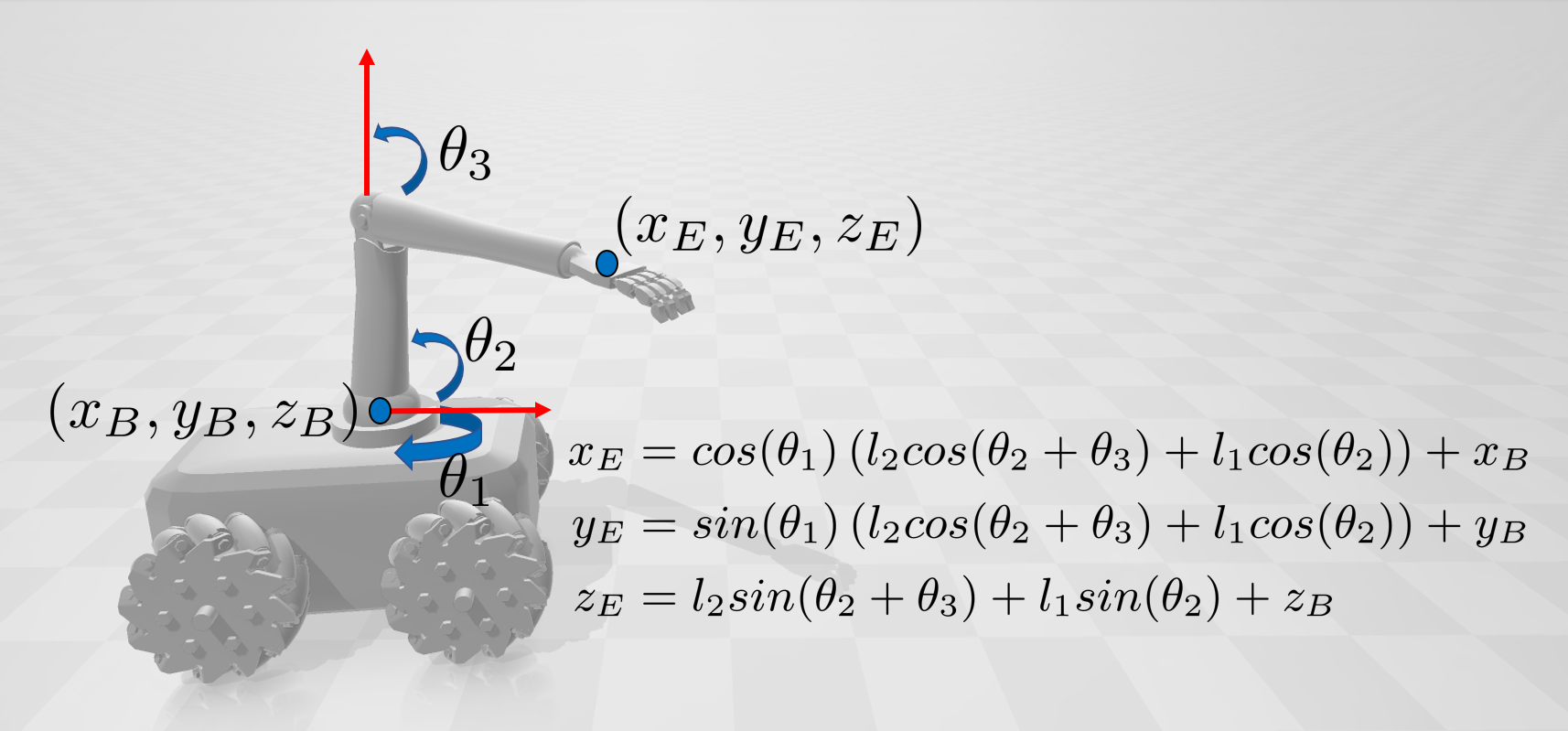}
    \caption{Example G- Mobile Robotic Arm Subjected to Joint and Base Uncertainties}
    \label{fig_G_Arm}
\end{figure}

The uncertain position of the end-effector of a mobile robotic arm is described in terms of the uncertain position of the base and the uncertain joint angles as shown in Figure \ref{fig_G_Arm}, \cite{Exa_Arm1, Exa_Arm2}. In this model, $\theta_i,i=1,2,3$ are the joint angles, $l_i,\ i=1,2$ are the length of the links, $(x_B,y_B,z_B)$ is the position of the base, and $(x_E,y_E,z_E)$ is the position of the end-effector of the mobile robotic arm. Uncertainty of the base and the joints are modeled as \begin{small}$x_B \sim \mathcal{U}niform(-0.1,0.1)$,
$y_B \sim \mathcal{N}(0,1)$, 
$z_B \sim \mathcal{B}eta(3,1)$, $\theta_1 \sim \mathcal{U}niform(-0.1,0.1)$, 
$\theta_2 \sim \mathcal{N}(\pi/4,1)$, $\theta_3 \sim \Gamma (1,2)$\end{small}.

Given the probability distributions of the uncertainties, we aim at finding the moments of the position of the end-effector of the mobile robotic arm. The position of the end-effector is a nonlinear function of the uncertainties. Hence, we can compute the moments in terms of the moments of the uncertainties as described in Section \ref{sec_UTrans}, e.g., $m_{x_E^{\alpha_1}y_E^{\alpha_2}z_E^{\alpha_3}}=\mathbb{E}[x_E^{\alpha_1}y_E^{\alpha_2}z_E^{\alpha_3}], \ (\alpha_1,\alpha_2,\alpha_3) \in \mathbb{N}^3$. For example, we obtain the moments of order 1 as \begin{small}$[m_{x_E},m_{y_E},m_{z_E}]=[0.34, 0.001, 1.43]$\end{small}, moments of order 2 as \begin{small}$[m_{x_E^2},m_{x_Ey_E},m_{x_Ez_E},m_{y_E^2},m_{y_Ez_E},m_{z_E^2}]=[1.12, 0.005, 0.35, 1.004, 0.001, 2.89]$\end{small}, and moments of order 3 as \begin{small}
$[m_{x_E^3},m_{x_E^2y_E},m_{x_E^2z_E},m_{xy_E^2},m_{x_Ey_Ez_E},m_{x_Ez_E^2},m_{y_E^3},m_{y_E^2z_E}, m_{y_Ez_E^2},m_{z_E^3}]=$\end{small} \begin{small}
$  [0.89,0.004,1.37,0.34,0.006,0.66,0.005,1.44, 0.003,6.16].$   
\end{small}One  can  verify  the  results  using the Monte Carlo simulation.


\section{Conclusion} \label{sec_con}
In this paper, we presented an exact moment based uncertainty propagation approach for nonlinear stochastic autonomous systems such as underwater, ground, and aerial vehicles, robotic  arms and walking robots. To this end, we introduced trigonometric and mixed-trigonometric-polynomial moments in the presence of arbitrary probabilistic uncertainties. We used such moments to  obtain deterministic linear moment-state dynamical  systems  to  describe the exact time evolution  of the  moments  of  the  uncertain states of nonlinear stochastic autonomous systems. As future work, we will use the proposed exact uncertainty propagation approach in nonlinear risk-aware motion planning, risk-aware nonlinear control, and nonlinear estimation problems. One can also use the obtained deterministic linear moment-state dynamical systems for stability analysis of stochastic nonlinear systems.


\end{document}